\documentclass{article}
\usepackage[utf8]{inputenc}
\usepackage{amsmath}
\usepackage{mathtools}
\usepackage{amsfonts}
\usepackage{physics}
\usepackage{float}
\usepackage{graphicx}
\usepackage{verbatim}
\usepackage{multirow}

\usepackage{bbm}
\usepackage{bm}
\usepackage{algorithmic}
\usepackage{algorithm}
\usepackage[round]{natbib}
\usepackage{amsthm} 
\usepackage{algorithmic}
\usepackage{algorithm}
\usepackage{geometry}
\usepackage{mleftright }
\usepackage{tabularx}
    \newcolumntype{L}{>{\raggedright\arraybackslash}X}
\usepackage{subcaption}

\makeatletter
\newcommand{\svast}{\bBigg@{3}}
\newcommand{\vast}{\bBigg@{4}}
\newcommand{\Vast}{\bBigg@{5}}
\makeatother

\begin{document}


\title{\bfseries \sffamily Efficient Bayesian inference for nonlinear state space models with univariate autoregressive state equation}
	\date{\small \today}
			\author{Alexander Kreuzer \footnote{Corresponding author: {E-mail: a.kreuzer@tum.de}}  \and Claudia Czado}
\date{%
	Zentrum Mathematik, Technische Universit\"at M\"unchen\\[2ex]%
	\today
}
			\maketitle
\vspace*{-0.2cm}

\begin{abstract}
Latent autoregressive processes are a popular choice to model time varying parameters. These models can be formulated as nonlinear state space models for which inference is not straightforward due to the high number of parameters. Therefore maximum likelihood methods are often infeasible and researchers rely on alternative techniques, such as Gibbs sampling. But conventional Gibbs samplers are often tailored to specific situations and suffer from high autocorrelation among repeated draws. We present a Gibbs sampler for general nonlinear state space models with an univariate autoregressive state equation. For this we employ an interweaving strategy and elliptical slice sampling to exploit the dependence implied by the autoregressive process. 
Within a simulation study we demonstrate the efficiency of the proposed sampler for bivariate dynamic copula models.
Further we are interested in modeling the volatility return relationship. Therefore we use the proposed sampler to estimate the parameters of stochastic volatility models with skew Student t errors and the parameters of a novel bivariate dynamic mixture copula model.
This model allows for dynamic asymmetric tail dependence.
Comparison to relevant benchmark models, such as the DCC-GARCH or a Student t copula model, with respect to predictive accuracy shows the superior performance of the proposed approach. 
\end{abstract}

\section{Introduction}

There are many situations where statistical models with constant parameters are no longer sufficient to appropriately represent certain aspects of the economy. For example it is well known that volatility of financial assets changes over time (\cite{schwert1989does}). This is why many models that allow for variation in the parameter have been proposed. There are time varying vector autoregressive models (\cite{primiceri2005time}, \cite{nakajima2011bayesian}), stochastic volatility models (\cite{kim1998stochastic}), GAM copula models (\cite{vatter2015generalized}, \cite{vatter2018generalized}) and many more. Stochastic volatility models and the bivariate dynamic copula model of \cite{almeida2012efficient} assume that the parameter follows a latent autoregressive process of order 1 (AR(1) process). These two models belong to the class of models that we will study.

In a general time varying parameter framework we consider a $d$ dimensional random variable  at time $t$, $\boldsymbol {Y_t} \in \mathbb{R}^d$, which is generated from a $d$ dimensional density $f(\cdot|s_t)$. We are interested in models, where the density $f(\cdot | s_t)$ has a univariate dynamic parameter $s_t \in \mathbb{R}$ following an AR(1) process. These models can be formulated as state space models with observation equation  
\begin{equation}
\begin{split}
\boldsymbol {Y_t}|s_t &\sim f(\boldsymbol {y_t}|s_t) \text{ independently},
\end{split}
\label{eq:obs}
\end{equation}
for $t=1, \ldots, T$. The state equation describes an AR(1) process with mean parameter $\mu \in \mathbb{R}$, persistence parameter $\phi \in (-1,1)$  and standard deviation parameter $\sigma \in (0,\infty)$ and is given by
\begin{equation}
s_t = \mu + \phi (s_{t-1} - \mu) + \sigma \epsilon_t,
\label{eq:states}
\end{equation}
where $\epsilon_t \sim N(0,1) \text{ iid}$ for $t=1, \ldots, T$ and $s_0|\mu,\phi,\sigma \sim N\left(\mu, \frac{\sigma^2}{1 - \phi^2}\right)$. In the state equation we assume Gaussian innovations $\epsilon_t$ but in the observation equation we do not put any restrictions on the density $f$. Thus we allow for nonlinear and non Gaussian state space models. Several established models can be analyzed within this framework.

By choosing $f(\cdot|s_t)$ as the univariate normal density with mean $0$ and variance $\exp({s_t})$, denoted by $\varphi(\cdot|0,\exp({s_t}{}))$, we obtain the stochastic volatility model (\cite{kim1998stochastic}) given by
\begin{equation}
\begin{split}
& Y_t|s_t \sim \varphi(y_t|0,\exp({s_t})) \text{ independently}, \\
& s_t = \mu + \phi (s_{t-1} - \mu) + \sigma \epsilon_t,
\end{split}
\label{eq:sv_model}
\end{equation}
for $t=1, \ldots, T$. By modeling the log variance as a latent AR(1) process this model allows for time varying volatility. Time varying volatility is a stylized fact of financial time series. In this context, the stochastic volatility model  has also shown better performance than the frequently used GARCH models (\cite{engle1982autoregressive},  \cite{bollerslev1986generalized}) for several data sets (\cite{yu2002forecasting}, \cite{chan2016modeling}). 
To allow for heavy tails and skewness, other distributions have been considered in the observation equation. One example is the stochastic volatility model with skew Student t errors (\cite{abanto2015bayesian}), which can also be analyzed within our framework.

Dependence modeling is another research area, where models that allow for time varying parameters have been introduced. Vine copulas (\cite{bedford2001probability}, \cite{aas2009pair}, \cite{czadoanalyzing}) are widely used models to capture complex dependence structures. To name a few, \cite{brechmann2013risk} and \cite{nagler2019model} employ vine copulas for forecasting the value at risk of a protfolio,  \cite{aas2016pair} gives an overview of applications of vine copulas in finance including asset pricing, credit risk management and portfolio optimization and  \cite{barthel2018dependence} model the association pattern between gap times with D-vine copulas to study asthma attacks.
A vine copula model is made up of different bivariate copulas with corresponding dependence parameters. Since dependencies may change over time, extensions that allow for variation in the dependence parameter have been proposed. \cite{vatter2015generalized} introduce a bivariate copula model, where the dependence parameter follows a generalized additive model. 
Another approach is the dynamic bivariate copula model proposed by \cite{almeida2012efficient} and \cite{hafner2012dynamic}, 
which we will analyze within our state space framework.
For this model, we consider single parameter copula families for which there is a one-to-one correspondence between the copula parameter, denoted by $\theta$, and Kendall's $\tau$. So we can express Kendall's $\tau$ as a function of the copula parameter $\theta$ and we write $\tau(\theta)$. The restriction of Kendall's $\tau$ to the interval $(-1,1)$ is removed by applying the Fisher' Z transformation $F_Z(x) = \frac{1}{2}\log(\frac{1+x}{1-x})$. This transformed time varying Kendall's $\tau$ is then modeled by an AR(1) process. More precisely, we consider $T$ bivariate random vectors, $(U_{t1}, U_{t2})_{t=1, \ldots, T} \in [0,1]^{T\times 2}$, corresponding to $T$ time points. We assume for $t=1, \ldots, T$ that
\begin{equation}
\begin{split}
&(U_{t1}, U_{t2})|\theta_t \sim c(u_{t1},u_{t2}; \theta_t) \text{ independently} \\
&s_t = \mu + \phi (s_{t-1} - \mu) + \sigma \epsilon_t, \text{ for }  s_t = F_Z(\tau(\theta_t))),
\end{split}
\label{eq:dcop}
\end{equation}
where $c(u_{t1},u_{t2}; \theta_t)$ is a bivariate copula density with parameter $\theta_t$. 

Nonlinear state space models as specified with \eqref{eq:obs} and \eqref{eq:states} are typically difficult to estimate, since there is a large number of parameters and likelihood evaluation requires high dimensional integration. This often makes maximum likelihood approaches infeasible. Gibbs sampling (\cite{geman1984stochastic}) is a frequently used Bayesian approach to infer parameters of such nonlinear state space models (\cite{carlin1992monte}). But the posterior samples, resulting from conventional Gibbs sampling, often suffer from high autocorrelation. Furthermore, the availability of the full conditional distributions or at least an efficient MCMC approach to sample from them is required. This is often tailored to specific situations. We present a Gibbs sampling approach that is designed to handle models with a latent AR(1) process and general likelihood functions as specified by the state space formulation in Equations \eqref{eq:obs} and \eqref{eq:states}. 
To sample from the associated posterior distribution we rely on elliptical slice sampling (\cite{murray2010elliptical}) and on an ancillarity-sufficiency interweaving strategy (\cite{yu2011center}). Elliptical slice sampling is used to sample the latent states. This allows us to exploit the Gaussian dependence structure, that is implied by the AR(1) process. But even if we provide efficient methods to sample from the full conditionals, the sampler may still suffer from the dependence among the parameters in the posterior distribution. Additionally, its performance may vary for different model parameterizations (\cite{fruhwirth2003bayesian}, \cite{strickland2008parameterisation}). This problem is tackled with the ancillarity-sufficiency interweaving strategy, where the parameters of the  latent AR(1) process are sampled from two different parameterizations. The decision between two parameterizations is avoided by using both.   This approach has already shown good results for several models, including univariate and multivariate stochastic volatility models (\cite{kastner2014ancillarity}, \cite{kastner2017efficient}).
The efficiency of our proposed sampler is illustrated with a simulation study. 

The second part of this work has a more applied focus and deals with modeling the volatility return relationship, i.e. the dependence  between an index and the corresponding volatility index. More precisely, we investigate the American index S$\&$P500 and its volatility index the VIX as well as the German index DAX and the VDAX.
 It is important to provide appropriate models for this relationship, since it has influence on hedging and risk management decisions (\cite{allen2012volatility}). 
 For our analysis we make use of a two step approach commonly used in copula modeling    (\cite{joe1996estimation}) motivated by Sklar's Theorem (\cite{Sklar}). We first model the marginal distribution with a univariate skew Student t stochastic volatility model. In the second step we model the dependence for which we propose a dynamic copula model allowing for asymmetric tail dependence. This model is a dynamic mixture of a Gumbel and a Student t copula and can be seen as an alternative to the symmetrized Joe-Clayton copula of \cite{patton2006modelling}.
Estimation is carried out through a two step approach, where we first estimate the marginal stochastic volatility models, fix their parameters at point estimates and then estimate the dynamic mixture copula. At both steps, estimation is straightforward with the proposed sampler.
Our model is able to capture several characteristics of the joint distribution of volatility and return. With respect to the marginal distribution we observe positive skewness for volatility indices compared to slight negative skewness for the return indices. In the dependence structure we identify asymmetry and time variation. Finally, we compare the proposed model to several restricted models with constant or symmetric dependence and to a bivariate DCC-GARCH model (\cite{engle2002dynamic}). Model comparison with respect to predictive accuracy shows the superiority of our approach.

To summarize, the main contribution of this paper is an approach to efficiently sample from the posterior distribution of general nonlinear state space models as specified by Equations \eqref{eq:obs} and \eqref{eq:states}. In addition, we propose a dynamic mixture copula for time varying asymmetric tail dependence. We discuss Bayesian inference for this model class and demonstrate how it can be utilized to model the volatility return relationship.

The outline of the paper is as follows: After the introduction, we discuss the proposed MCMC approach in Section \ref{sec:mcmc}. In Section \ref{sec:illustration} we investigate the efficiency of the sampler for bivariate dynamic copula models through an extensive simulation study. Section \ref{sec:application} deals with modeling the volatility return relationship and Section \ref{sec:concluion} concludes.

\section{Bayesian inference}
\label{sec:mcmc}

In the following, we denote an observation of the $d$-dimensional random vector $\boldsymbol {Y_t}$  by $\boldsymbol {y_t}$ and the associated data matrix is given by $Y=(\boldsymbol {y_1}, \ldots, \boldsymbol {y_T})^\top \in \mathbb{R}^{T \times d}$.
To subset vectors and matrices we make use of the following notation:
For sets of indices $A$ and $B$ we set
$\boldsymbol x_A = (x_i)_{i \in A}$ for a vector $\boldsymbol x$ and $X_{A;B} = (x_{ij})_{i \in A, j \in B}$ for a matrix $X$. We use a capital letter to refer to a matrix and small letters to refer to its components. The set $\{n,\ldots, k\}$ of integers will be abbreviated by $n:k$.

We consider the state space model as specified by Equations \eqref{eq:obs} and \eqref{eq:states}.
To obtain a fully specified Bayesian model we equip the parameters $\mu$, $\phi$ and $\sigma$ with prior distributions. We follow \cite{kastner2014ancillarity}, who propose the following prior distributions for the latent AR(1) process of the stochastic volatility model:
\begin{equation}
\mu \sim N(0,\sigma_{\mu}^2), ~~\frac{\phi+1}{2} \sim Beta(a_{\phi}, b_{\phi}) ,~~ \sigma^2 \sim Gamma\left(\frac{1}{2}, \frac{1}{2B_{\sigma}}\right),
\label{eq:priors}
\end{equation}
where $\sigma_{\mu}, a_{\phi}, b_{\phi}, B_{\sigma} >0$. Our standard choice for the prior hyperparameters is $\sigma_{\mu}=100,~ a_{\phi} = 5,~ b_{\phi} = 1.5$ and $ B_{\sigma}=1$ as in \cite{kastner2016dealing}.
With these prior distributions our Bayesian model is complete.  
For sampling from the posterior distribution of this model we should take into account that sampling efficiency may highly depend on the model parameterization (\cite{fruhwirth2003bayesian}, \cite{strickland2008parameterisation}). \cite{yu2011center} differentiate between two parameterizations: A sufficient augmentation and an ancillary augmentation scheme. In our case a sufficient augmentation is characterized by an observation equation that is free of the parameters $\mu, \phi$ and $\sigma$ and only depends on the latent states $\boldsymbol {s_{1:T}}$. In this case $\boldsymbol {s_{1:T}}$ is a sufficient statistics for the parameters $\mu, \phi $ and $\sigma$. In an ancillary augmentation the state equation is independent of the parameters $\mu, \phi $ and $\sigma$, then $\boldsymbol {s_{1:T}}$ is an ancillary statistics for the parameters $\mu, \phi$ and $\sigma$.
 The standard parameterization of our model is already a sufficient augmentation and we refer to this parameterization as given by Equations \eqref{eq:obs} and \eqref{eq:states} as (SA).
\begin{equation*}
(SA): \hspace*{0.3cm}
\quad
\!
\begin{aligned}
& \boldsymbol {Y_t}|s_t \sim f(\boldsymbol {y_t}|s_t) \text{ independently}\\
& s_t = \mu + \phi (s_{t-1} - \mu) + \sigma \epsilon_t.
\end{aligned}
\end{equation*}
  An ancillary augmentation is obtained by the following parameterization
\begin{equation}
\tilde s_t = \frac{s_t - \mu -\phi(s_{t-1} - \mu)}{\sigma}, \text{ with inverse } s_t = \mu + \phi(s_{t-1} - \mu) + \sigma \tilde s_t,
\label{eq:rec}
\end{equation}
for $t=1 ,\ldots, T$. This reparameterization is obtained by solving Equation \eqref{eq:states} for $\epsilon_t$ and implies that the state space model is given by
\begin{equation*}
(AA): \hspace*{0.3cm}
\quad
\!
\begin{aligned}
&\boldsymbol {Y_t}|\boldsymbol{\tilde  s_{1:T}},s_0, \mu, \phi, \sigma \sim f(\boldsymbol {y_t}|s_t(\boldsymbol{\tilde  s_{1:T}},s_0, \mu, \phi, \sigma))\text{ independently}\\
&\tilde s_t \sim N(0,1) \text{ independently},
\end{aligned}
\end{equation*}
where $s_t(\boldsymbol{ \tilde s_{1:T}},s_0, \mu, \phi, \sigma)$ is the function that calculates $ s_t$ according to \eqref{eq:rec}.  We refer to this model representation as (AA). Instead of deciding between (SA) and (AA), we combine them in an ancillarity-sufficiency interweaving strategy (\cite{yu2011center}), given by
\begin{itemize}
\item a)  Sample  $\boldsymbol {s_{0:T}}$ in (SA) from
$
\boldsymbol {s_{0:T}}|Y,  \mu, \phi, \sigma  
$
.
\item b) Sample $(\mu, \phi, \sigma)$ in (SA) from 
$
\mu, \phi, \sigma|Y,\boldsymbol {s_{0:T}}
$.
\item c) Move to (AA) via
$
\tilde s_t = \frac{s_t - \mu -\phi(s_{t-1} - \mu)}{\sigma},
$
for $t=1, \ldots, T$.
\item d) Sample $(\mu, \phi, \sigma)$ in (AA) from 
$
\mu, \phi, \sigma|Y, s_0,\boldsymbol{ \tilde s_{1:T}}
$ .
\item e) Move back to (SA) via the recursion
$
s_t = \mu + \phi(s_{t-1} - \mu) + \sigma \tilde s_t
$
for $t=1, \ldots, T$.
\end{itemize}

\cite{kastner2014ancillarity} employed interweaving for the stochastic volatility model and showed its superior performance with an extensive simulation study. For the stochastic volatility model, \cite{kastner2014ancillarity} propose to  move between a sufficient augmentation and a reparameterization of the latent states $\boldsymbol {s_{1:T}}$  given by $ s^K_t = \frac{s_t - \mu}{\sigma}$. Within this reparameterization parameters can be sampled conveniently from its full conditional distribution by recognizing a linear regression model. This is possible for the standard stochastic volatility model, but not in our case since our sampler is designed to handle more general likelihood functions. Therefore we have chosen the reparameterization of (SA) such that it is optimal in the sense of \cite{yu2011center}, i.e. we move between a sufficient and a ancillary augmentation.

Note that reducing the sampler to the first two steps a) and b) results in a standard Gibbs sampler in (SA). This sampler typically suffers from the dependence among the parameters $\mu, \phi, \sigma$ and the latent states $\boldsymbol {s_{1:T}}$ in the posterior distribution.

\subsubsection*{Step a: Sampling of the latent states in the sufficient augmentation }
\label{sec:latstat}
To sample the latent states $\boldsymbol {s_{0:T}}$ from its full conditional in (SA) we make use of elliptical slice sampling as proposed by \cite{murray2010elliptical}. It was developed  for models, where dependencies are generated through a latent multivariate normal distribution.
In (SA), the AR(1) structure implies, that the vector $\boldsymbol {s_{0:T}}|\mu, \phi, \sigma$ has a $(T+1)$ dimensional multivariate normal distribution with mean vector $\boldsymbol{\mu^{AR}}$ and covariance matrix $\Sigma^{AR}$ given by
\begin{equation}
\boldsymbol{\mu^{AR}} = \left( {\begin{array}{c}
\mu \\
\mu \\
\vdots \\
\mu
\end{array} } \right)
\in \mathbb{R}^{T+1}, ~~~
\Sigma^{AR} = \frac{\sigma^2}{1-\phi^2}
  \left( {\begin{array}{ccccc}
   1 & \phi & \phi^2 & \ldots & \phi^T  \\
   \phi & 1 & \phi & \ldots & \phi^{T-1} \\
            \vdots & \vdots & \vdots &  & \vdots \\
                        \phi^T & \phi^{T-1} & \phi^{T-2} & \ldots & 1 \\
  \end{array} } \right)
   \in \mathbb{R}^{(T+1) \times (T+1)}.
\label{eq:vecmat}
\end{equation}
(see e.g. \cite{brockwell2002introduction}, Chapter 2.2).
The posterior density is proportional to
\begin{equation*}
\left( \prod_{t=1}^T f(\boldsymbol {y_t}|s_t) \right)  \varphi(\boldsymbol {s_{0:T}}|\boldsymbol {\mu^{AR}}, \Sigma^{AR}) \pi(\mu) \pi(\phi) \pi(\sigma), 
\end{equation*}
where $\boldsymbol {\mu^{AR}}$ and $\Sigma^{AR}$ are given in \eqref{eq:vecmat} and $\pi(\cdot)$ denotes the corresponding prior density as specified in \eqref{eq:priors}.
The initial state can be sampled from its full conditional density given by
\begin{equation*}
f(s_0|\boldsymbol {s_{1:T}}, \mu, \phi, \sigma) = \varphi(s_0| \mu+\phi(s_1-\mu), \sigma^2).
\end{equation*}

The full conditional density of the latent states $\boldsymbol {s_{1:T}}$ is given by
\begin{equation*}
f(\boldsymbol {s_{1:T}}|Y, s_0, \mu, \phi, \sigma) \propto 
\left(\prod_{t=1}^T f(\boldsymbol {y_t}|s_t)\right) \varphi(\boldsymbol {s_{1:T}}|\boldsymbol {\mu_{1:T|0}}, \Sigma_{1:T|0}), 
\end{equation*}
with a corresponding mean vector $\boldsymbol {\mu_{1:T|0}}$ and covariance matrix $\Sigma_{1:T|0}$. The mean vector and the covariance matrix of the conditional distribution are derived in a more general way in Appendix A.
By reparameterizing the model with
$
\boldsymbol {s_{1:T}'} = \boldsymbol {s_{1:T}} - \boldsymbol {\mu_{1:T|0}},
$
we impose a multivariate normal prior with zero mean. We obtain the situation elliptical slice sampling was designed for. However updating the whole $T$ dimensional vector $\boldsymbol {s_{1:T}}$ with elliptical slice sampling at once will lead to high autocorrelation in the posterior draws. This is illustrated in Section \ref{sec:illustration} and was also observed by \cite{hahn2019efficient}, where elliptical slice sampling was used for linear regression models. \cite{hahn2019efficient} circumvent this problem by partitioning the vector $\boldsymbol {s_{1:T}}$ into smaller blocks. 
We follow this approach and partition the set $\{1, \ldots, T\}$ into $m$ different blocks $B_1, \ldots, B_m \subset \{1, \ldots, T\}$. Let $a_i = \min_{s \in B_i} s$ denote the minimal and $b_i = \max_{s \in B_i} s$ denote the maximal index in the $i$-th block. The blocks are chosen such that  $B_i = \{t \in \{1, \ldots, T\}: a_i \leq t \leq b_i \}$, for $i = 1, \ldots, m$.
The full conditional density for the $i$-th block can be expressed as
$
f(\boldsymbol {s_{B_i}}|Y, s_0, \boldsymbol {s_{-B_i}}, \mu, \phi, \sigma) \propto 
\left(\prod_{t \in B_i} f(\boldsymbol {y_t}|s_t)\right) f(\boldsymbol {s_{B_i}}|s_0, \boldsymbol {s_{-B_i}}, \mu, \phi, \sigma), 
$
where $-B_i = \{1, \ldots, T\} \setminus B_i$. The vector $\boldsymbol {s_{B_i}}|s_0, \boldsymbol {s_{-B_i}}, \mu, \phi, \sigma$ is multivariate normal distributed with mean denoted by $\boldsymbol {\mu_{B_i|}}$ and covariance matrix $\Sigma_{B_i|}$ (see Appendix A) and therefore the full conditional density can be written as
\begin{equation*}
f(\boldsymbol {s_{B_i}}|Y, s_0, \boldsymbol {s_{-B_i}}, \mu, \phi, \sigma) \propto 
\left(\prod_{t \in B_i} f(\boldsymbol {y_t}|s_t)\right) \varphi(\boldsymbol {s_{B_i}}|\boldsymbol {\mu_{B_i|}}, \Sigma_{B_i|}).
\end{equation*}
To sample the latent states of the $i$-th block $\boldsymbol {s_{B_i}}$ from its full conditional we proceed as follows
\begin{itemize}
\item Set
$
\boldsymbol {s'_{B_i}} = \boldsymbol {s_{B_i}} - \boldsymbol {\mu_{B_i|}}
$
\item Draw $\boldsymbol {s'_{B_i}}$ from the density 
\begin{equation*}
f(\boldsymbol {s'_{B_i}}|Y, s_0, \boldsymbol {s_{-B_i}}, \mu, \phi, \sigma) \propto 
\left(\prod_{t \in B_i} f(\boldsymbol {y_t}|s_t)\right) \varphi(\boldsymbol {s'_{B_i}}|\boldsymbol{0}, \Sigma_{B_i|}), 
\end{equation*}
using elliptical slice sampling, where $\varphi(\boldsymbol {s'_{B_i}}| \boldsymbol 0,\Sigma_{B_i|})$ is interpreted as the prior density for $\boldsymbol {s'_{B_i}}$.

\item Set
$
\boldsymbol {s_{B_i}} = \boldsymbol {s'_{B_i}} + \boldsymbol {\mu_{B_i|}}
$ 
\end{itemize}

\subsubsection*{Step b: Sampling of the constant parameters in the sufficient augmentation}
\label{sec:SA}
In (SA) the observation equation only depends on $\boldsymbol {s_{1:T}}$ and is independent of the parameters $\mu$, $\phi$ and $\sigma$. The parameters $\mu$, $\phi$ and $\sigma$ only depend on $\boldsymbol {s_{0:T}}$. This allows to use the same approach  as in \cite{kastner2014ancillarity} to sample the parameters $\mu, \phi$ and $\sigma$ in (SA). We reparameterize the model such that proposals can be found using Bayesian linear regression. We define $\gamma = \mu (1-\phi) $ and the state equation is given by
\begin{equation*}
s_t = \gamma + \phi s_{t-1} + \sigma\eta_t, 
\end{equation*}
where $\eta_t \sim N(0,1)$. For fixed $\boldsymbol {s_{0:T}}$, this is a linear regression model with regression parameters $\gamma, \phi$ and variance $\sigma^2$. Proposals for $(\mu, \phi, \sigma)$ are found and accepted or rejected as described in \cite{kastner2014ancillarity} Section 2.4 (two block sampler).

\subsubsection*{Step d: Sampling of the constant parameters in the ancillary augmentation}
\label{sec:AA}
To sample $\mu, \phi$ and $\sigma$ in (AA) we deploy a random walk Metropolis-Hastings scheme with Gaussian proposal, where the proposal variance or covariance matrix is adapted during the burn-in period. For the adaptions we use the Robbins Monro process (\cite{robbins1985stochastic}) as suggested by \cite{garthwaite2016adaptive}. More details are given in Appendix A.

\subsubsection*{Implementation}
For the implementation of the sampler we use $\texttt{Rcpp}$ (\cite{eddelbuettel2011rcpp}) which allows to embed C++ code into \texttt{R}. In addition we make use of \texttt{rvinecopulib} (\cite{nagler2018rvinecopulib}) to evaluate copula densities and of \texttt{RcppEigen} (\cite{bates2013fast}). For sampling $(\mu, \phi, \sigma)$ in (SA) we use corresponding parts of the implementation of the \texttt{R} package $\texttt{stochvol}$ (\cite{kastner2016dealing}). The \texttt{R} package \texttt{coda} (\cite{plummer2008coda}) is used to compute effective sample sizes in the following section.

\section{Illustration of the proposed sampler for bivariate dynamic copula models}
\label{sec:illustration}

We illustrate the MCMC sampler we proposed in the previous section for the bivariate dynamic copula model of \cite{almeida2012efficient}.
\cite{kastner2014ancillarity} have already shown that interweaving improves sampling efficiency a lot for the stochastic volatility model.
We investigate if this is also the case for the bivariate dynamic copula model.
Further we study how the sampling efficiency is affected by the chosen block size and by the data generating process (DGP).
Therefore we perform an extensive simulation study. 
We consider different modifications of the sampler. A sampler is specified by a vector $(\text{b,i})$ which indicates its blocksize (b) and if interweaving is used (i=I) or not (i=NI). We consider ten different sampler specifications $(\text{b,i}) \in \{1,5,20,100,T\} \times \{\text{I}, \text{NI} \}$, where $T$ is the length of the time series. By using blocks of size $T$, we obtain the sampler which updates the parameters $\boldsymbol {s_{1:T}}$ jointly with elliptical slice sampling. If we turn off interweaving (i=NI) we obtain a standard Gibbs sampler updating parameters in the sufficient augmentation. These samplers are run for different simulated data sets. A data set is simulated from the bivariate dynamic copula model (see \eqref{eq:dcop}) with parameters: Family, $T, \mu, \phi, \sigma$. The parameters are chosen from the following grid  
(Family, $T, \mu, \phi, \sigma$) $\in \{\text{Gauss}, \text{eClayton}\}$ $\times \{500, 1000, 1500\} \times \{0,1\} \times \{0,0.1,0.5,0.9,0.99\} \times \{0.05,0.1,0.2\}$.
Here, eClayton denotes the extended Clayton copula, which extends the Clayton copula to allow for negative Kendall's $\tau$ values. More precisely, the extended Clayton copula 
 has the following density
\begin{equation*}
   c(u_1,u_2;\theta) =
   \begin{cases}
    c_{Clayton}(u_1, u_2; \theta) \text{ if } \theta \geq 0 \\
     c_{Clayton}(1-u_1, u_2; -\theta) \text{ if } \theta < 0,  \\
   \end{cases}
\end{equation*}
where $c_{Clayton}(\cdot, \cdot;\theta)$ is the density of the bivariate Clayton copula with parameter $\theta$ (see \cite{joe2014dependence}, Chapter 4).
So the extended Clayton copula density is equal to the Clayton copula density for non negative Kendall's $\tau$ and equal to a 90 degree rotation of the Clayton copula density for negative Kendall's $\tau$.
Among the different DGPs, most distinct values are considered for  $\phi$. We expect its choice to be influential since it controls the dependence among the latent variables. 
With this grid we obtain 180 different DGPs. For each of the different DGPs we generate 100 simulated data sets and for each data set we run the 10 different samplers with the correctly specified copula family for 25000 iterations and discard the first 5000 iterations for burn-in. So, in total we obtain 18000 simulated data sets and each of the 10 samplers is run 18000 times.

\begin{figure}[H]
\centering
\includegraphics[trim={0 7cm 0 0},width=1\textwidth]{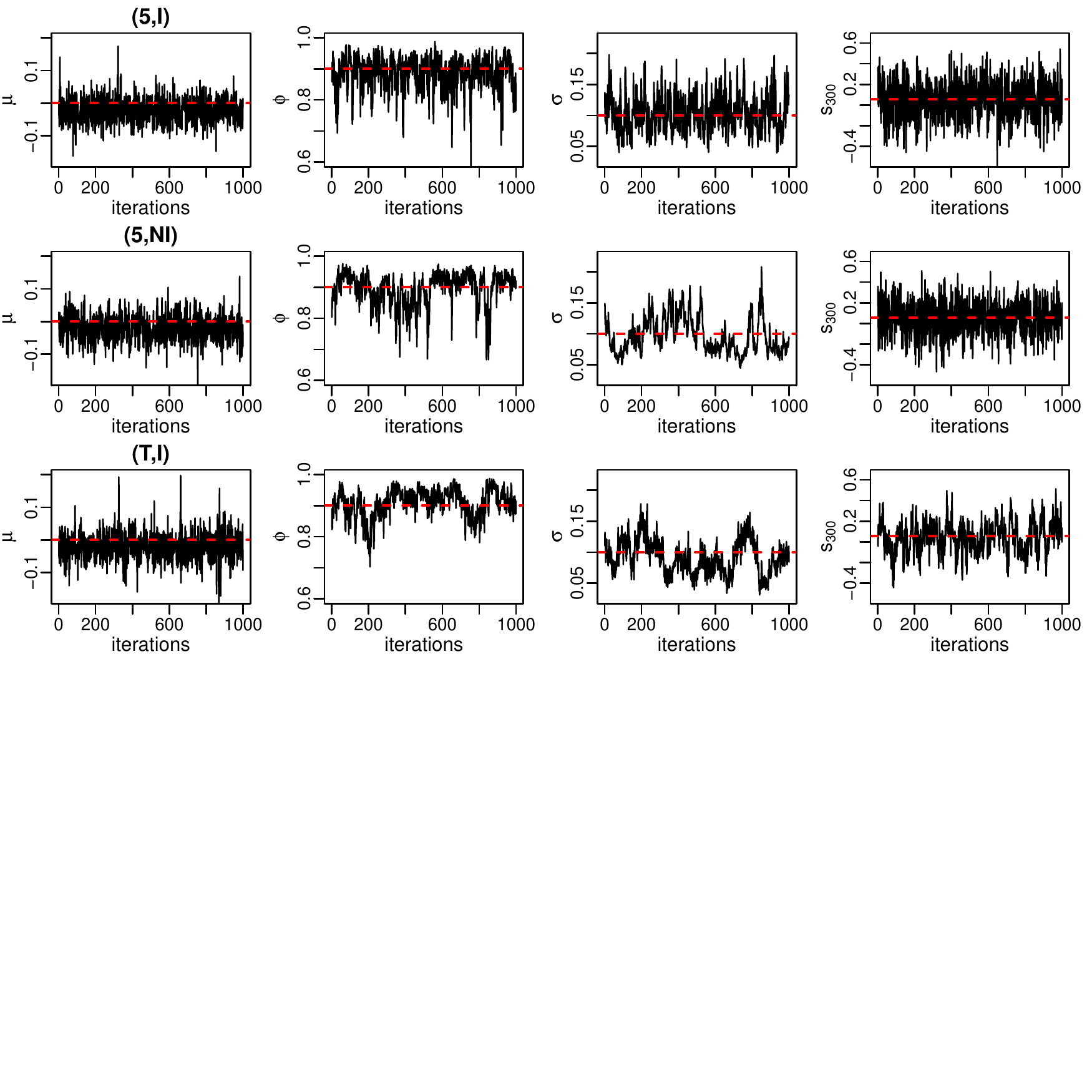}
\caption{Trace plots of 1000 MCMC draws based on a total of 25000 iterations, where the first 5000 draws are discarded for burn-in and the remaining 20000 draws are thinned with factor 20. The trace plots  are shown for the parameters $\mu, \phi$, $\sigma$ and $s_{300}$ for three different sampler specifications: (5,I) (top row), (5,NI) (middle row), (T,I) (bottom row). The corresponding data was generated from the following DGP: $T=1000$, Family=eClayton, $\mu=0$, $\phi=0.9$, $\sigma=0.1$. True values are added in red (dashed).}
\label{fig:trace_sim_rev}
\end{figure}

Figure \ref{fig:trace_sim_rev} shows trace plots of different parameters ($\mu$, $\phi$, $\sigma$, $s_{300}$) based on one simulated data set for three different sampler specifications. We consider specification (5,I), and the same specification with interweaving turned off, i.e. (5,NI) and the specification (T,I).
We observe that all three samplers produce posterior samples covering the true values. Further, the trace plots suggest that the (5,I) sampler achieves better mixing than the other two considered samplers.

The runtime of the sampler is mainly affected by the choice of $T$ and the sampler specification. From Table \ref{tab:runtime} we see that the runtime is increasing in $T$ and that interweaving adds considerable additional runtime. 

\begin{table}[H]
\footnotesize
\centering
\begin{tabular}{l|rrrrr|rrrrr}
& (1,I) & (5,I) & (20,I) & (100,I) & (T,I) & (1,NI) & (5,NI) & (20,NI) & (100,NI) & (T,NI) \\ 
  \hline
$T=500$ & 0.6 & 0.5 & 0.5 & 0.7 & 0.8 & 0.3 & 0.3 & 0.3 & 0.4 & 0.5 \\ 
$T=1000$ & 1.1 & 1.0 & 1.1 & 1.3 & 1.6 & 0.7 & 0.5 & 0.6 & 0.8 & 1.1 \\ 
$T=1500$ & 1.7 & 1.4 & 1.6 & 1.9 & 2.6 & 1.0 & 0.7 & 0.9 & 1.2 & 1.8 \\ 
   \hline
\end{tabular}
\caption{Average runtime in minutes for 25000 draws. We consider averages for different sampler specifications and different values of $T$. The sampler was run on a Linux cluster with CPU Intel Xeon E5-2697 v3.}
\label{tab:runtime}
\end{table}

To measure efficiency of the samplers we consider the effective sample size per minute which we call effective sampling rate, similar to \cite{hosszejni2019approaches}. We average the effective sampling rate of 100 runs, where the same sampler and the same DGP  was used. Then we obtain 180 average effective sampling rates (AESR) per sampler.
Since the AESR decreases for higher values of $T$ for every sampler, we compare the ten different samplers among DGPs with the same value for $T$.  For a fixed $T$ we have 60 AESR values per sampler. For each sampler, the minimum of these 60 values (mAESR) is given in Table \ref{tab:bc_simsum}.
We consider the minimum since we are interested in samplers that are reliable for all DGPs. 
We see that for the parameters $\mu$, $\phi$ and $\sigma$ sampler specifications with interweaving have higher mAESR values, while for the latent states specifications without interweaving perform better.
Although interweaving adds additional runtime, it still increases the mAESR of $\mu$, $\phi$ and $\sigma$ considerably.
Further we observe that choosing the blocksize too big results in very low mAESR values for the latent states. In this case many parameters are updated jointly with elliptical slice sampling which results in high autocorrelation among consecutive draws. The performance of samplers with blocksize $T$ is especially poor. For these samplers the length of the time series ($T=$ 500, 1000, 1500) also has strong effects. For example the mAESR for $\mu$ for specification (T,NI) decreases by 85 $\%$ from 78 to 11 when the length of the time series is increased from 500 to 1500. 
The most inefficient sampler is (T,NI), the sampler without interweaving and with the largest blocksize.
In our opinion the best results are obtained for
sampler specification (5,I). It provides the highest mAESR values for $\mu$, $\phi$ and $\sigma$ for all choices of $T$ and also provides rather high mAESR values for the latent states.

\begin{table}[H]
\footnotesize
\centering
\begin{tabular}{r|rrrrr|rrrrr}
& (1,I) & (5,I) & (20,I) & (100,I) & (T,I) & (1,NI) & (5,NI) & (20,NI) & (100,NI) & (T,NI) \\ 
  \hline
\multicolumn{10}{c}{$T=500$}\\
$\mu$ & 2829 & 3178 & 1813 & 642 & 381 & 437 & 838 & 780 & 277 & 78 \\ 
$\phi$ & 471 & 749 & 556 & 191 & 60 & 266 & 326 & 232 & 78 & 24 \\ 
$\sigma$ & 272 & 411 & 263 & 58 & 33 & 53 & 74 & 56 & 23 & 6 \\ 
$s$(a) & 938 & 2982 & 2484 & 199 & 36 & 1347 & 4942 & 3908 & 273 & 49 \\ 
$s$(m) & 346 & 1092 & 441 & 35 & 5 & 496 & 1011 & 627 & 41 & 5 \\ 
    \hline
\multicolumn{10}{c}{$T=1000$}\\
$\mu$ & 1167 & 1365 & 840 & 309 & 180 & 172 & 330 & 371 & 121 & 21 \\ 
$\phi$ & 202 & 269 & 212 & 73 & 17 & 81 & 97 & 76 & 27 & 6 \\ 
$\sigma$ & 97 & 175 & 133 & 25 & 9 & 18 & 25 & 21 & 12 & 1 \\ 
$s$(a) & 298 & 1011 & 1312 & 99 & 12 & 400 & 1911 & 1982 & 133 & 16 \\ 
$s$(m) & 115 & 417 & 200 & 15 & 1 & 161 & 559 & 296 & 17 & 2 \\ 
   \hline
   \multicolumn{10}{c}{$T=1500$}\\
$\mu$ & 711 & 833 & 578 & 200 & 111 & 89 & 218 & 231 & 75 & 11 \\ 
$\phi$ & 120 & 145 & 107 & 43 & 9 & 38 & 49 & 42 & 14 & 3 \\ 
$\sigma$ & 63 & 109 & 84 & 16 & 3 & 10 & 14 & 13 & 8 & 1 \\ 
$s$(a) & 162 & 583 & 879 & 66 & 7 & 247 & 1168 & 1359 & 90 & 10 \\ 
$s$(m) & 61 & 239 & 112 & 9 & 1 & 95 & 426 & 176 & 11 & 1 \\ 
\hline
\end{tabular}
\caption{For different lengths of the time series $(T=500,1000,1500)$ the minimum of 60 AESR values (mAESR) corresponding to 60 different DGPs is shown for different parameters and ten different sampler specifications. In the $s(a)$ row we calculate the mAESR based on the average of the effective sampling rates of $s_0, \ldots, s_T$, while in the $s(m)$ row the mAESR values are calculated based on the minimum of the effective sampling rates of $s_0, \ldots, s_T$.}
\label{tab:bc_simsum}
\end{table}

In addition to the previous analysis, we investigate how different DGPs affect the samplers. Therefore we consider the best sampler according to Table \ref{tab:bc_simsum}, i.e. sampler specification (5,I). In addition we have a look at the same sampler specification without interweaving (5,NI) and the same sampler with joint updates of the latent states, i.e. (T,I). We consider the AESR for $\sigma$, which is usually the parameter which causes most problems.
Table \ref{tab:minmaxAESR} shows for each of these three samplers the DGPs with $T=1000$ which resulted in the lowest and the highest AESR for $\sigma$.
For the (5,I) specification satisfactory AESR values, ranging from 175 to 456, are obtained for all DGPs.
 In the (T,I) specification the latent states are updated jointly. In scenarios with strong dependence among the latent states ($\phi$ = 0.99) this sampler performs poorly whereas the best performance of the sampler was seen for a DGP with low dependence among the latent states ($\phi$=0). The (5,NI) specification is a standard Gibbs sampler in the sufficient augmentation. We see that for this specification the AESR values vary a lot, ranging from 25 to 558. The best performance of this sampler specification was seen for a DGP with high persistence ($\phi=0.99$). \cite{kastner2014ancillarity}  studied different sampling schemes for the stochastic volatility model and have also seen that a standard Gibbs sampler in the sufficient augmentation performs well for scenarios with high persistence.

\begin{table}[ht]
\footnotesize
\centering
\begin{tabular}{lr|rr|rr|rr}

&&\multicolumn{2}{c}{(5,I)}&\multicolumn{2}{c}{(T,I)}&\multicolumn{2}{c}{(5,NI)}\\
  \hline
 \multirow{2}{*}{AESR($\sigma$)}&&min &max&min &max&min &max \\
&&175 &456&9 &121 &25 &558 \\
\hline
 \multirow{2}{*}{DGP}&family & Gauss & eClayton& eClayton & eClayton& Gauss & eClayton\\
&$\mu$ & 0 & 1& 0 & 0& 0 & 0\\
&$\phi$ & 0.1 & 0.9& 0.99 & 0& 0.1 & 0.99\\
&$\sigma$ & 0.2 & 0.2& 0.2 & 0.05& 0.05 & 0.2\\
   \hline
\end{tabular}
\caption{For three sampler specifications ((5,I), (T,I), (5,NI)) we show the DGPs with $T=1000$ which resulted in the highest and in the lowest AESR values for $\sigma$, respectively. The corresponding AESR is also shown.}
\label{tab:minmaxAESR}
\end{table}

Lastly we compare our sampler to the coarse grid sampler employed by \cite{almeida2012efficient}. We already covered six DGPs that were also analyzed by \cite{almeida2012efficient}. Instead of running their sampler we make the comparison with respect to these six cases. There are several points which make the comparison slightly less reliable. First \cite{almeida2012efficient} did not report exact computation times but they note that 100 000 iterations of their sampler take about 15 minutes. We use this number to calculate the AESR values from the effective sample sizes they report in their paper. Second their calculations were performed on a different computer and third they use different prior distributions for $\phi$ and $\sigma^2$. But we think that this comparison should still give us a rough idea of how the sampling efficiencies compare to each other. From Table \ref{tab:compCG} we see that the (5,I) specification considerably outperforms the coarse grid sampler (CG). For every parameter ($\mu$, $\phi$, $\sigma$) we obtain way higher AESR values.

\begin{table}[H]
\footnotesize
\centering
\begin{tabular}{rrrr|rr|rr|rr}
  \multicolumn{4}{c}{DGP} & \multicolumn{2}{c}{AESR($\mu$)}& \multicolumn{2}{c}{AESR($\phi$)}& \multicolumn{2}{c}{AESR($\sigma$)} \\
family&$\mu$ & $\phi$ & $\sigma$ & (5,I) & CG & (5,I) & CG & (5,I
) & CG \\ 
  \hline
Gauss&$1$ & $0.9$ & $0.1$ &  8187 & 2130 & 527 & 81 & 393 & 50 \\ 
Gauss&$1$ & $0.1$ & $0.2$  & 1637 & 305 & 364 & 83 & 437 & 69 \\ 
Gauss&$0$ & $0.9$ & $0.2$ & 9935 & 4150 & 549 & 150 & 316 & 85 \\ 
eClayton&$1$ & $0.9$ & $0.1$ & 8382 & 2088 & 539 & 74 & 397 & 46 \\ 
eClayton&$1$ & $0.1$ & $0.2$ & 1778 & 279 & 378 & 71 & 416 & 52 \\ 
eClayton&$0$ & $0.9$ & $0.2$ & 10079 & 4375 & 574 & 175 & 323 & 97 \\ 
   \hline
\end{tabular}
\caption{Comparison of sampler specification (5,I) and the coarse grid sampler (CG) of \cite{almeida2012efficient} for six different DGPs with $T=1000$ with respect to the AESR of $\mu$, $\phi$ and $\sigma$.}
\label{tab:compCG}
\end{table}

\section{Application: Modeling the volatility return relationship}
\label{sec:application}

We investigate the volatility return relationship through the bivariate joint distribution of a stock index and the corresponding volatility index. 
The joint distribution of return and volatility incorporates all the marginal information as well as information about the dependence, which are both relevant for hedging and risk management (\cite{allen2012volatility}). Since there has already been evidence for asymmetry in the joint distribution of volatility and return (\cite{allen2012volatility}, \cite{fink2017regime}) models which are able to handle these characteristics are necessary.

The two step copula modeling approach motivated by Sklar's theorem (\cite{Sklar}) provides a very flexible method for the construction of multivariate distributions. We can combine arbitrary marginal distributions with any copula. Here, we propose a bivariate model that combines the skew Student t stochastic volatility model (\cite{abanto2015bayesian}) for the margins with a novel dynamic mixture copula.  This model allows for asymmetry and heavy tails in the marginal distribution as well as for time varying asymmetric tail dependence in the dependence structure. Both, the marginal as well as the copula model can be estimated with the proposed sampler.

\subsection*{Marginal model}
The stochastic volatility model with skew Student t errors is obtained by replacing the normal distribution of the stochastic volatility model (\cite{kim1998stochastic})  by a skew Student t distribution. This allows for heavy tails and skewness. 
The stochastic volatility model with skew Student t errors as considered by \cite{abanto2015bayesian} is given by
\begin{equation}
\begin{split}
&Y_t = \exp(\frac{s_t}{2}) \epsilon_t  \\
&s_t = \mu + \phi (s_{t-1} - \mu) + \sigma \epsilon_t, 
\end{split}
\label{eq:stsv}
\end{equation}
where $\epsilon_t|\alpha,df \sim sst(\epsilon_t|\alpha, df)$ independently for $t=1, \ldots, T$. We denote by $sst(\epsilon_t|\alpha, df)$ the density of the standardized skew Student t distribution with parameters $\alpha \in \mathbb{R}$ and $df>2$ (cf. Appendix B).
Compared to our framework this model has two additional parameters $\alpha$ and $df$. For these additional parameters we choose the following prior distributions
\begin{equation}
\alpha \sim N(0,100), ~~~~~~ df \sim N_{>2}(5,25),
\label{eq:stsv_priors}
\end{equation}
where $N_{>2}$ denotes the normal distribution truncated to $(2, \infty)$. We need to ensure that $df > 2$ since the standardized skew Student t distribution would not be well defined otherwise.

 Conditional on $\alpha$ and $df$ our sampler can be applied directly to sample $(\mu, \phi, \sigma, \boldsymbol {s_{0:T}})$ from its full conditional. Another approach, which lead to better mixing, is to include the parameters $\alpha$ and $df$ in the interweaving strategy. The sampler is slightly modified in the following way:
\begin{itemize}
\item a)  Sample $\boldsymbol {s_{0:T}}$ from
$
\boldsymbol {s_{0:T}}|Y,  \mu, \phi, \sigma, \alpha, df  
$
.
\item b) Sample $(\mu, \phi, \sigma, \alpha, df)$ in (SA) from 
$
\mu, \phi, \sigma, \alpha, df|Y,\boldsymbol {s_{0:T}}
$ .
\item c) Move to (AA) via
$
\tilde s_t = \frac{s_t - \mu -\phi(s_{t-1} - \mu)}{\sigma},
$
for $t=1, \ldots, T$.
\item d) Sample $(\mu, \phi, \sigma, \alpha, df)$ in (AA) from 
$
\mu, \phi, \sigma, \alpha, df|Y, s_0,\boldsymbol{\tilde s_{1:T}}
$.
\item e) Move back to (SA) via the recursion
$
s_t = \mu + \phi(s_{t-1} - \mu) + \sigma \tilde s_t
$
for $t=1, \ldots, T$.
\end{itemize}

For step a) we proceed as described in Section \ref{sec:mcmc}. 
For step b) we draw $\alpha$ and $df$ from its univariate full conditional distributions using Metropolis-Hastings, similar to Step d) in Section \ref{sec:mcmc}.
The parameters $(\mu, \phi, \sigma)$ are drawn from its full conditional as described in Section \ref{sec:mcmc}. 
 For step d) we investigated different blocking strategies for the parameters $(\mu, \phi, \sigma, \alpha, df)$. We compared the different strategies with respect to effective sample sizes and decided to use the following three blocks: $(\mu,df), (\phi, \sigma)$ and $\alpha$. Each block is updated using Metropolis-Hastings as in Step d) in Section \ref{sec:mcmc}.

\subsection*{Dependence model}
Dependence among financial assets is often modeled with a Student t copula. 
This copula allows for tail dependence symmetric in the upper and lower tail. Evidence against the assumption of symmetric tail dependence has been provided and models to handle this characteristic have become necessary (\cite{patton2006modelling}, \cite{nikoloulopoulos2012vine}, \cite{jondeau2016asymmetry}). \cite{patton2006modelling} proposes the symmetrized Joe-Clayton  copula. This is a modification of the BB7 copula (\cite{joe2014dependence}, Chapter 4) that is symmetric if upper and lower tail dependence coincide, which he describes as a desirable property.
In the application of \cite{nikoloulopoulos2012vine} the Student t copula provides the best fit in terms of the likelihood. But they argue that if the focus is on the tails a BB1 or BB7 copula might be more appropriate. The BB1 and BB7 copulas have two parameters which might not be enough, if we want to model three characteristics in a flexible way: upper tail dependence, lower tail dependence and overall dependence as measured with Kendall's $\tau$.  
 \ We provide another approach to relax the symmetric tail dependence assumption. We propose a mixture of a Student t and a extended Gumbel copula with parameters $\tau \in (-1,1)$, $ \nu >2$ and $ p \in [0,1]$ given by
\begin{equation}
C^M(u_1,u_2; \tau, \nu, p) =  p C^t(u_1,u_2;\tau,\nu) + (1-p) C^G(u_1,u_2;\tau))  ,
\label{eq:mixcop}
\end{equation}
where  $C^t$ is the bivariate Student t copula specified by Kendall's $\tau$ and the degree of freedom $\nu$ and $C^G$ is the bivariate extended Gumbel copula specified by Kendall's $\tau$. The extended Gumbel copula is defined similarly to the extended Clayton copula in Section \ref{sec:mcmc}, i.e. its density is equal to the Gumbel copula density for positive values of Kendall's $\tau$ and equal to a 90 degree rotation of the Gumbel copula density for negative Kendall's $\tau$ values. Both copulas $C^t$ and $C^G$ share the dependence parameter $\tau$ and we expect the mixture copula to have a similar strength of dependence. 
The corresponding Kendall's $\tau$ of the mixture copula is given by
\begin{equation}
\begin{split}
\tau^M
=&\int_{(0,1)^2} C^M(u_1,u_2;\tau, \nu, p) c^M(u_1,u_2;\tau, \nu, p) du_1du_2 =(p^2 + (1-p)^2) \tau + \\
&+\int_{(0,1)^2} p(1-p) \left( C^G(u_1,u_2;\tau)) (c^t(u_1,u_2;\tau,\nu) + C^t(u_1,u_2;\tau,\nu) c^G(u_1,u_2;\tau))\right) du_1du_2. \\
\end{split}
\label{eq:taumix}
\end{equation}

We evaluated the integral in \eqref{eq:taumix}  numerically for different values of $\tau$, $p$ and $\nu$ and observed only negligible difference between $\tau$ and $\tau^M$. The upper and lower tail dependence coefficients $\lambda_M^L$ and $\lambda_M^U$ of the mixture copula can, for $\tau>0$, be obtained as
\begin{equation*}
\begin{split}
\lambda_M^L(\tau, p, \nu) =& \lim_{u\rightarrow 0} \frac{C^M(u,u)}{u} = \lim_{u\rightarrow 0} \frac{p C^t(u_1,u_2;\tau,\nu) + (1-p) C^G(u_1,u_2;\tau)}{u}\\
=& p 2 T_{\nu+1}(-\sqrt{\frac{(\nu+1)(1-\sin(\pi  \frac{\tau}{2}))}{1+\sin(\pi  \frac{\tau}{2})}})  + 0\\
\lambda_M^U(\tau, p, \nu) =   & p 2 T_{\nu+1}(-\sqrt{\frac{(\nu+1)(1-\sin(\pi  \frac{\tau}{2}))}{1+\sin(\pi  \frac{\tau}{2})}})  + (1-p)(2 - 2^{1-\tau}),
\end{split}
\end{equation*}
\noindent
where we used the well known formulas for the tail dependence coefficients of the Student t and the Gumbel copula (\cite{joe2014dependence}, Chapter 4).
Whereas the upper and lower tail dependence coefficients measure dependence in the upper right and lower left corner, we are also interested in the dependence in the upper left and the lower right corner when $\tau<0$.
We consider the following tail dependence coefficients in the upper left corner $\lambda_M^{UL}$ and in the lower right corner $\lambda_M^{LR}$ if $\tau<0$
\begin{equation*}
\begin{split}
\lambda_M^{LR} &= \lambda_M^L(-\tau, p, \nu) , ~~~
\lambda_M^{UL} = \lambda_M^U(-\tau, p, \nu), 
\end{split}
\end{equation*}
analogous to the definition of quarter tail dependence in \cite{fink2017regime}.

The tail dependence coefficient of the mixture copula is a linear combination of the tail dependence coefficients of its two components, the Student t and the Gumbel copula. The Student t copula has symmetric tail dependence, whereas the Gumbel copula has upper but no lower tail dependence. So we expect upper tail dependence to be higher than lower tail dependence in the mixture copula. The amount of asymmetry in the tails is controlled by $p$, whereas the copula is symmetric in the tails for $p=1$ and the level of asymmetry increases as we decrease $p$.
So this copula allows for great flexibility: The overall dependence can be described by Kendalls's $\tau$, the degrees of freedom parameter controls the upper and lower tail dependence coefficient and $p$ controls the difference between upper and lower tail dependence.  This is visualized in Figure \ref{fig:taildep} in Appendix C.
Note that the desirable property according to \cite{patton2006modelling} of symmetry in case of coinciding upper and lower tail dependence is here fulfilled.  If we expected higher lower than upper tail dependence we can replace the Gumbel copula by a survival Gumbel copula which has the density $c^{SG}(u_1,u_2) = c^G(1-u_1,1-u_2)$. 
To allow for time variation we use the mixture copula $C^M$ of \eqref{eq:mixcop} within the dynamic bivariate copula model of \cite{almeida2012efficient}. A nonlinear state space model for $T$ bivariate random vectors $(U_{t1}, U_{t2})_{t=1, \ldots, T} \in [0,1]^{T\times 2}$, corresponding to $T$ time points, is given by
\begin{equation}
\begin{split}
&(U_{t1}, U_{t2}) \sim c^M(u_{t1}, u_{t2}; \tau_t, \nu, p) \text{ independently} \\
&s_t = \mu + \phi (s_{t-1} - \mu) + \sigma \epsilon_t, ~~ s_t = F_Z(\tau_t)
\end{split}
\label{eq:dynmix}
\end{equation}
for $t=1, \ldots, T$.
We assign a uniform prior on $[0,1]$ for $p$, a normal prior with mean 5 and standard deviation 20 truncated to the interval $(2,\infty]$ for $\nu$ and the same priors as in \eqref{eq:priors} for the remaining parameters. Sampling is done in the following way.
\begin{itemize}
\item Draw $\log(\frac{p}{1-p})$ and $\log(\nu -2 )$ from its univariate full conditionals with random walk Metropolis-Hastings with Gaussian proposal (proposal standard deviation: 0.3).
\item Draw $\mu, \phi, \sigma, \boldsymbol{s_{0:T}}$ conditioned on $p$ and $\nu$ as in Section \ref{sec:mcmc}.
\end{itemize}

\subsection*{Two step estimation}
We consider the S$\&$P500 (SPX)  and its volatility index the VIX as well as the DAX and its volatility index the VDAX. The daily log returns from 2006 to 2013 of these indices are obtained from Yahoo finance (https://finance.yahoo.com). With approximately 250 trading days per year this results in 2063 observations, visualized in Figure \ref{fig:dlogret} in Appendix D. The corresponding data matrix with 2063 rows and 4 columns is denoted by $Y$.

Combining the marginal and the dependence model we obtain that for $T$ bivariate random vectors $(Y_{t1},Y_{t2})_{t=1, \ldots, T} \in \mathbb{R}^{T \times 2}$ the following holds

\begin{equation}
\begin{split}
&(Y_{t1},Y_{t2}) \sim \\
&C^M\left(ssT\left( \frac{y_{t1}}{{\exp(s_{t1}^{st}/2)}} \Bigg | \alpha_{1}^{st}, {df}_{1}^{st}\right), ssT\left( \frac{y_{t2}}{{\exp(s_{t2}^{st}/2)}}\Bigg | \alpha_{2}^{st}, {df}_{2}^{st}\right);F_Z^{-1}( s_{t}^{cop}),  \nu^{cop},  p^{cop}     \right)
\end{split}
\label{eq:fmod}
\end{equation}

\noindent
independently, where
\begin{equation*}
\begin{split}
s_{tj}^{st} &= \mu_{j}^{st} + \phi_{j}^{st} (s_{t-1;j}^{st} - \mu_{j}^{st}) + \sigma_j^{st} \epsilon_{tj}^{st} \\
s_{t}^{cop} &= \mu^{cop} + \phi^{cop} (s_{t-1}^{cop} - \mu^{cop}) + \sigma^{cop}\epsilon_{t}^{cop}
\end{split}
\end{equation*}
\noindent
and $\epsilon_{tj}^{st}, \epsilon_{t}^{cop} \sim N(0,1)$ iid, $\alpha_{j}^{st}, df_{j}^{st}$ as in \eqref{eq:stsv_priors}, $\nu^{cop}, p^{cop}$ as in \eqref{eq:dynmix} and $\mu_{j}^{st}$, $\mu^{cop}$, $\phi_{j}^{st}$, $\phi^{cop}$, $\sigma_{j}^{st}$, $\sigma^{cop}$, $s_{0j}^{st}$, $s_{0}^{cop}$ as in \eqref{eq:states} and \eqref{eq:priors} for $j=1,2$ and $t=1, \ldots T$. Here, $ssT$ denotes the distribution function of the standardized skew Student t distribution (cf. Appendix B). We refer to the probability integral transforms $ssT\left( {y_{t1}}{{\exp(-s_{t1}^{st}/2)}} \big| \alpha_{1}^{st}, {df}_{1}^{st}\right)$ and  $ssT\left( {y_{t2}}{{\exp(-s_{t2}^{st}/2)}}\Big| \alpha_{2}^{st}, {df}_{2}^{st}\right)$ for $t=1, \ldots, T$ as copula data.

For inference we rely on a two step approach. We first estimate marginal distributions and based on the resulting estimated copula data we estimate the copula parameters. This approach is also called inference for margins (\cite{joe1996estimation}) and is commonly used in (Bayesian) copula modeling (\cite{min2011bayesian}, \cite{almeida2012efficient}, \cite{smith2015copula}, \cite{gruber2015sequential}, \cite{loaiza2018time}).

\begin{figure}[H]
\centerline{%
\includegraphics[trim={0 11.25cm 0 0},width=1.0\textwidth]{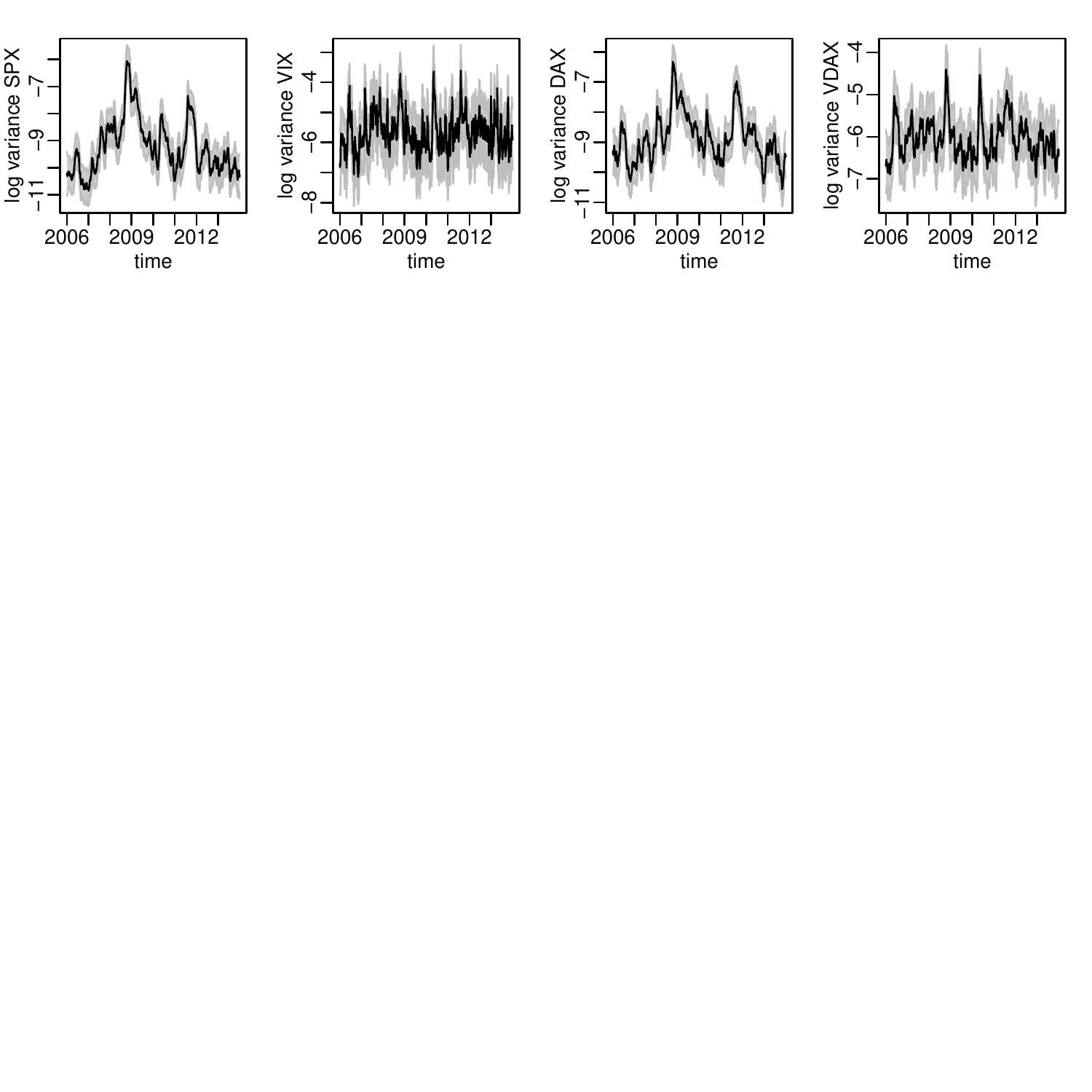}%
}%
\caption{Posterior mode estimates of the daily log variances of the four skew Student t stochastic volatility models for the indices SPX, VIX, DAX, VDAX from 2006 to 2013 plotted against time. The $90 \%$ credible region, added in grey, is constructed from the $5\%$ and $95\%$ posterior quantiles.}
\label{fig:vola}
\end{figure}

First we fit a skew Student t stochastic volatility model for each of the indices. For each index we run the sampler (5,I) for 31000 iterations and discard the first 1000 draws as burn-in.
As it is typical for financial data all indices show a high persistance parameter $\phi$ (Posterior mode estimates for $\phi$: SPX: $0.99$, VIX: $0.90$, DAX: $0.99$, VDAX: $0.96$). A notable difference is that for stock indices we observe negative skewness, whereas for the volatility indices positive skewness is observed (Posterior mode estimates for $\alpha$: SPX: $-0.51$, VIX: $1.33$, DAX: $-0.48$, VDAX: $0.96$). Evidence for negative skewness has also been observed for the log returns of other stock indices, as e.g., for the NASDAQ by \cite{abanto2015bayesian}.  Posterior mode estimates, posterior quantiles and effective samples sizes for several parameters of the four marginal models are summarized in Table \ref{tab:marginal} in Appendix D. The estimated daily log variances are shown in Figure \ref{fig:vola}. In the end of 2009, the estimated variances are high for all indices due to the financial crisis.

In the next step we obtain data on the [0,1] scale by applying the probability integral transform using the posterior mode estimates of the marginal parameters. We refer to this data as pseudo copula data and it is obtained as
\begin{equation*}
\hat u_{tj} = ssT\left(y_{tj}\exp(-\frac{\hat s_{tj}^{st}}{2});\hat \alpha_j^{st}, \hat {df}_j^{st}\right), 
\end{equation*} 
where $\hat s_{tj}^{st}, \hat \alpha_j^{st}, \hat{ df}_j^{st}$ are the posterior mode estimates of the corresponding marginal skew Student t stochastic volatility model for $t=1, \ldots, T, j=1,\ldots, 4$. 
In the copula data marginal characteristics are removed and what is left is information about the dependence structure.
Based on the pseudo copula data two dynamic mixture copula models are fitted, one corresponding to the pair (SPX,VIX) and one corresponding to the pair (DAX,VDAX). For each pair we obtain 31000 iterations with the sampler specification (5,I) and discard the first 1000 draws as burn-in.  The posterior mode estimates for $p$ are $0.29$ for the model for (SPX,VIX) and $0.66$ for the model for (DAX,VDAX), respectively. (Further, posterior statistics for the model parameters $\mu, \phi, \sigma, p$ and $\nu$ are shown in Table \ref{tab:cop} in Appendix D). So both fitted models allow for asymmetric tail dependence, whereas the asymmetry is stronger for the (SPX,VIX) model. For these models tail dependence in the upper left corner $\lambda_M^{UL}$ is stronger than the one in the lower right corner $\lambda_M^{LR}$. This means that joint extreme comovements, where the stock index decreases and the volatility index increases are more likely to occur than vice versa, which agrees with the statement that the market reacts more extreme in bad market situations (\cite{sun2018leverage}).

\begin{figure}[H]
\centerline{%
\includegraphics[trim={0 6cm 0 0},width=0.9\textwidth]{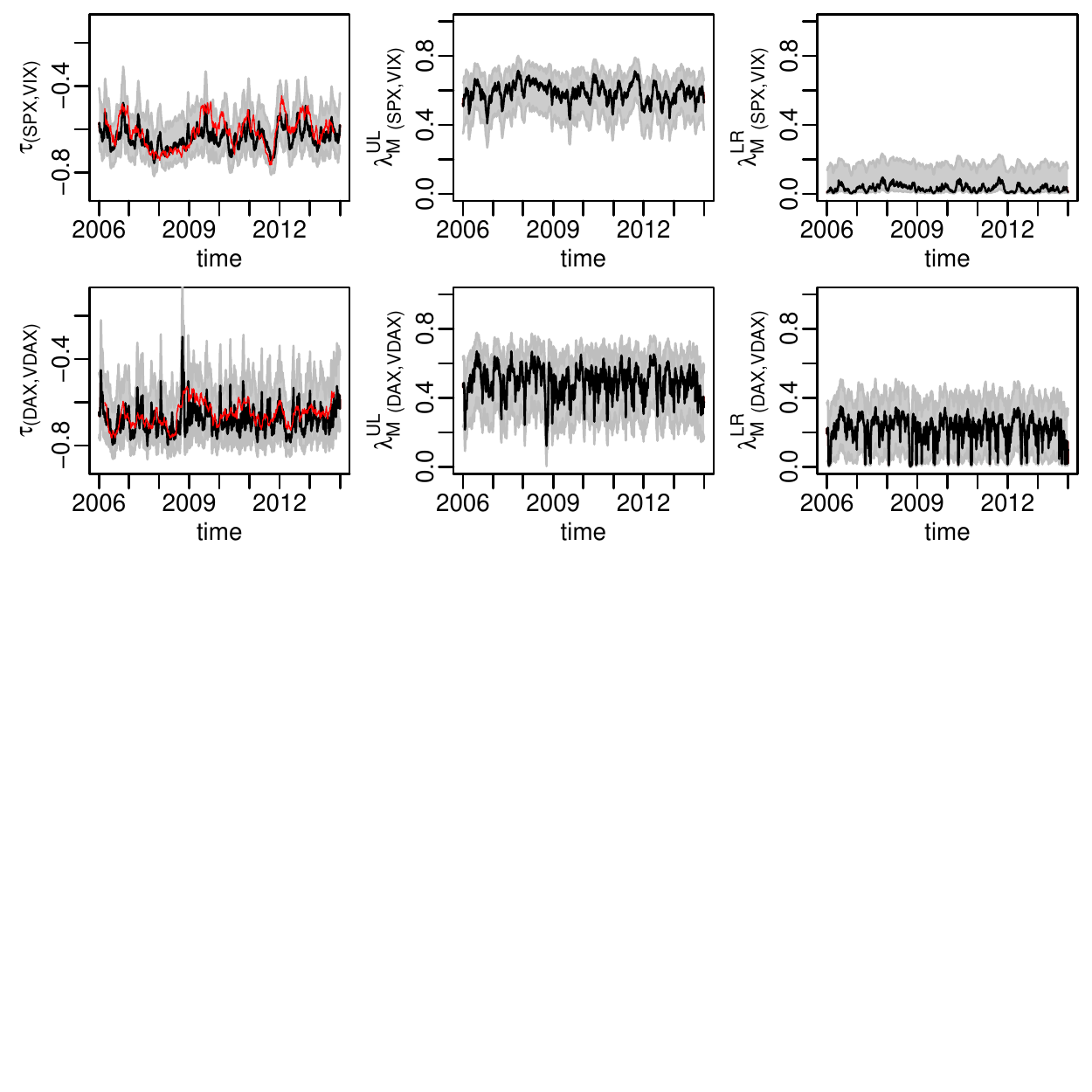}%
}%
\caption{In these plots the top row corresponds to the pair (SPX,VIX), the bottom row to the pair (DAX,VDAX).  The first column  shows posterior mode estimates of Kendall's $\tau$, where the Kendall's $\tau$ obtained from rolling window estimates (Kendall's $\tau$ at time $t$ is estimated as empirical Kendall's $\tau$ based on the 50 observations before and after time $t$) is added in red. The middle column shows posterior mode estimates of $\lambda_M^{UL}$ and the right column shows estimates of $\lambda_M^{LR}$. Credible regions, constructed from the $5\%$ and $95\%$ posterior quantiles, are added in grey.}
\label{fig:dyntautaildep}
\end{figure}

The time varying estimates for Kendall's $\tau$ and the tail dependence coefficients are shown in Figure \ref{fig:dyntautaildep}. The figure visualizes the asymmetry in tail dependence. We also observe changes in tail dependence as time evolves: for (SPX,VIX), $\lambda_M^{UL}$ ranges from $0.41$ to $0.71$ and $\lambda_M^{LR}$ from $0.004$ to $0.10$ . For the pair (DAX,VDAX), $\lambda_M^{UL}$ ranges from $0.12$ to $0.67$  and $\lambda_M^{LR}$ from $0.003$ to $0.35$. This variation over time in tail dependence goes hand in hand with variation in Kendall's $\tau$. For (SPX,VIX), Kendall's $\tau$ ranges from $-0.76$ to $-0.48$ and for (DAX,VDAX) from  $-0.80$ to $-0.30$.

\subsection*{Out of sample predictions}
We aim to further support the findings we obtained through the dynamic copula model, i.e. that the dependence structure is asymmetric and varies over time. Therefore we consider several restrictions with respect to the dependence structure. Giving up time variation leads a constant mixture copula, giving up asymmetry leads to a dynamic Student t copula and giving up time variation and asymmetry leads to a constant Student t copula. 
In addition, we compare our model to the frequently used dynamic conditional correlation (DCC) GARCH model of \cite{engle2002dynamic}. The DCC-GARCH allows for time varying symmetric dependence.
So we take five different models into consideration. These models are summarized in Table \ref{tab:model_spec}. The model $\mathcal{M}_{dyn}^{mix}$ is the model given in \eqref{eq:fmod}, which was also used for the previous analysis.

We compare the models with respect to cumulative pseudo log predictive scores (\cite{kastner2019sparse}), which are obtained by evaluating the corresponding density at point estimates, instead of averaging over all posterior draws.
 In comparison to other multivariate scoring rules, such as the energy score or the variogram score (\cite{scheuerer2015variogram}), pseudo log predictive scores have here the advantage that they can be computed fast since they require only one evaluation of the density per observation. 
We consider $T+K$ observations of dimension two, stored in the data matrix $Y_{1:(T+K);1:2}$, where the first $T$ observations are used to train the model and the last $K$ are used for testing.

\begin{table}[H]
\centering
\begin{tabular}{l|l|r|rrrr}
\hline
model& specification & \multicolumn{1}{c}{margin} & \multicolumn{2}{c}{dependence}\\

   &  & asymmetric & asymmetric & dynamic \\ 
  \hline
 $\mathcal{M}_{dyn}^{mix}$ & sstSV +  dynamic mixture copula & Yes & Yes & Yes \\ 
   \hline
    $\mathcal{M}_{const}^{mix}$ & sstSV +  constant mixture copula & Yes & Yes & No \\ 
     \hline
     $\mathcal{M}_{dyn}^{t}$ & sstSV +  dynamic Student t copula & Yes & No & Yes\\ 
      \hline
      $\mathcal{M}_{const}^{t}$ & sstSV +  constant Student t copula & Yes& No & No\\ 
       \hline      
      $\mathcal{M}_{DCC}$ & DCC(1,1)-GARCH(1,1) & No & No & Yes \\ 
       \hline

\end{tabular}
\caption{Different models considered for comparison. Models are specified by: marginal model + copula model. The skew Student t stochastic volatility model as given in \eqref{eq:stsv} is denoted by sstSV. The mixture copula is defined in \eqref{eq:mixcop}. If a copula is dynamic, the corresponding copula is considered within the dynamic bivariate copula model framework of \cite{almeida2012efficient} given in \eqref{eq:dcop}.} 
\label{tab:model_spec}
\end{table}

The DCC-GARCH model is estimated based on $T$ data points in the training period. Based on this model, we obtain rolling one-day ahead estimates of the covariance matrix for each day in the test set. The pseudo log predictive scores are obtained  by  evaluating the corresponding multivariate normal log densities at the observations. To estimate the DCC-GARCH models we used the R package $\texttt{rmgarch}$ of \cite{ghalanos2012rmgarch}.

Similarly, the model $\mathcal{M}_{dyn}^{mix}$ is estimated with the training data. Instead of computing daily updates for all model parameters we fix the constant parameters at their posterior mode estimates to save computation time. The dynamic parameters are updated daily and the one-day ahead forecasts are obtained by evolving the AR(1) process. To obtain the pseudo log predictive scores we evaluate the density implied by \eqref{eq:fmod} at the corresponding observations. Appendix D contains a detailed description of this procedure. The pseudo log predictive scores for $\mathcal{M}_{const}^{mix}$, $\mathcal{M}_{dyn}^{t}$ and $\mathcal{M}_{const}^{t}$ are obtained similarly.

This procedure for calculating the pseudo log predictive scores is applied for both data sets corresponding to the pairs (SPX,VIX) and (DAX,VDAX). Using the last two years (2012 - 2013) of our data set as test data yields $K=517$. As training period we use $T=1000$ which corresponds to a training period of approximately four years. 

Table \ref{tab:lp} summarizes the cumulative pseudo log predictive scores.
In both cases, for the (SPX,VIX) as well as for the (DAX,VDAX) data, the best model is provided by the dynamic mixture copula model $\mathcal{M}_{dyn}^{mix}$. 
Furthermore, we see that in both cases the constant mixture copula model $\mathcal{M}_{const}^{mix}$ is preferred over the constant and dynamic Student t copula models $\mathcal{M}_{const}^{t}$ and $\mathcal{M}_{dyn}^{t}$. For the (DAX,VDAX) data, the second best model is provided by the constant mixture copula model $\mathcal{M}_{const}^{mix}$. For this data the rolling window estimates of Kendall's $\tau$ in Figure \ref{fig:dyntautaildep} vary less than for the (SPX,VIX) data. For the (SPX,VIX) data, the DCC-GARCH $\mathcal{M}_{DCC}$ yields the second best cumulative pseudo log predictive scores.

\begin{table}[ht]
\centering
\begin{tabular}{llllllll}
 & $\mathcal{M}_{dyn}^{mix}$  & $\mathcal{M}_{const}^{mix}$ &  $\mathcal{M}_{dyn}^{t}$ & $\mathcal{M}_{const}^{t}$ & $\mathcal{M}_{DCC}$\\ 
  \hline
(SPX,VIX) & \textbf{2740.7} & 2733.5 & 2732.0 & 2725.9 & 2737.2 \\ 
(DAX,VDAX) &  \textbf{2817.7} & 2814.1 & 2810.1 & 2809.9 & 2789.0 \\

   \hline
\end{tabular}
\caption{Cumulative pseudo log predictive scores for the models $\mathcal{M}_{dyn}^{mix}$, $\mathcal{M}_{const}^{mix}$, $\mathcal{M}_{dyn}^{t}$, $\mathcal{M}_{const}^{t}$ and  $\mathcal{M}_{DCC}$.}
\label{tab:lp}
\end{table}

\section{Conclusion}
\label{sec:concluion}

We propose a sampler, applicable to general nonlinear state space models with univariate autoregressive state equation. Sampling efficiency is demonstrated for bivariate dynamic copula models within a simulation study. Furthermore we use the sampler to estimate the parameters of a dynamic bivariate mixture copula model. This mixture copula model turns out to be a good candidate to model the volatility return relationship, since in our application it produces more accurate forecasts than a bivariate DCC-GARCH model or a Student t copula model.

In this work, there are two objectives that might be extended: The sampler and the bivariate mixture copula model. The sampler could be extended  to allow for a broader class of models. For example we might consider autoregressive processes of higher order in the state equation. In this case we can still rely on elliptical slice sampling and on an interweaving strategy.  Another extension could relax the assumption of a Gaussian dependence structure in the state equation by replacing the  autoregressive process by  a D-vine copula model. In this case elliptical slice sampling can no longer be applied to sample the latent states and an alternative sampling method is required.

The bivariate dynamic mixture copula could serve as a building block for regular vine copula models. Thus we could extend the bivariate model to arbitrary dimensions. This is interesting if we study not only the bivariate  volatility return relationship, but for example the dependence structure among several exchange rates.

\bibliographystyle{spbasic}    
\bibliography{References}{}

\newpage
\section*{Appendix A. Details on the sampling procedure}

\subsection*{Sampling of the latent states in the sufficient augmentation}

Here we derive $\boldsymbol {\mu_{B_i|}}$ and $\Sigma_{B_i|}$. By the conditional independence assumptions of the AR(1) process and the way we defined the blocks $B_1, \ldots, B_m$ we obtain 
\begin{equation*}
\begin{split}
f(\boldsymbol {s_{B_i}}|s_0, \boldsymbol {s_{-B_i}}, \mu, \phi, \sigma) &= f(\boldsymbol {s_{B_i}}|  s_{a_i-1},s_{b_i+1}, \mu, \phi, \sigma), \text{ for } i = 1, \ldots, m-1, \text{ and }\\
f(\boldsymbol {s_{B_m}}|s_0, \boldsymbol {s_{-B_m}}, \mu, \phi, \sigma) &= f(\boldsymbol {s_{B_m}}|  s_{a_m-1}, \mu, \phi, \sigma).
\end{split}
\end{equation*}
Conditional on $\mu, \phi$ and $\sigma$, the vector $(s_{a_i-1}, \boldsymbol {s_{B_i}}, s_{b_i+1})$ is multivariate normal distributed with mean vector $\boldsymbol {\mu^{AR}_{(a_i-1, {B_i}, b_i+1)}} \in \mathbb{R}^{c_i+2}$ and covariance matrix $ \Sigma^{AR}_{(a_i-1, {B_i}, b_i+1);(a_i-1, {B_i}, b_i+1)} \in \mathbb{R}^{(c_i+2)\times(c_i+2)}$, where $c_i$ is the cardinality of 
 $B_i$.
Thus the vector
$\boldsymbol {s_{B_i}}|s_{a_i-1}, s_{b_i+1}, \mu, \phi, \sigma$ follows a multivariate normal distribution with mean vector $\boldsymbol {\mu_{B_i|}}$ and covariance matrix $\Sigma_{B_i|}$ given by
\begin{equation}
\begin{split}
\boldsymbol {\mu_{B_i|}}&= \boldsymbol{
\mu^{AR}_{B_i}} + \Sigma^{AR}_{B_i;(a_i-1,b_i+1)}  \frac{1-\phi^2}{(1-\phi^{2(c_i+1)})\sigma^2}   \left( {\begin{array}{cc}
1 & -\phi^{c_i+1}\\
-\phi^{c_i+1} & 1\\
\end{array} } \right)
\left( {\begin{array}{c}
s_{a_i-1} - \mu \\
s_{b_i+1} - \mu \\
\end{array} } \right), \\
\Sigma_{B_i|}&=\Sigma^{AR}_{B_i;B_i} -\Sigma^{AR}_{B_i;(a_i-1,b_i+1)} \frac{1-\phi^2}{(1-\phi^{2(c_i+1)})\sigma^2}  
\left( {\begin{array}{cc}
1 & -\phi^{c_i+1}\\
-\phi^{c_i+1} & 1\\
\end{array} } \right)
\Sigma^{AR}_{(a_i-1,b_i+1);B_i}.
\label{eq:condmeanvar}
\end{split}
\end{equation}

The vector $\boldsymbol {s_{B_m}}|s_{a_m-1},  \mu, \phi, \sigma$ corresponding to the last block is multivariate normal distributed with mean vector $\boldsymbol {\mu_{B_m|}}$ and covariance matrix $\Sigma_{B_m|}$ obtained as
\begin{equation*}
\begin{split}
\boldsymbol {\mu_{B_m|}} &= \boldsymbol{
\mu^{AR}_{B_m}} + \Sigma^{AR}_{B_m;a_m-1}  \frac{1-\phi^2}{\sigma^2} (s_{a_m-1} - \mu), \\
\Sigma_{B_m|} &=\Sigma^{AR}_{B_m;B_m} -\Sigma^{AR}_{B_m;a_m-1} \frac{1-\phi^2}{\sigma^2}  
\Sigma^{AR}_{a_m-1;B_m}.
\end{split}
\end{equation*}

We need to sample from $N(\boldsymbol 0, \Sigma_{B_i|})$ several times during elliptical slice sampling. Instead of working with the $c_i \times c_i$ covariance matrix $\Sigma_{B_i|}$ we can more efficiently sample from the $c_i$ dimensional normal distribution  by using the conditional independence assumptions of the AR(1) process. It holds that

\begin{equation*}
\begin{split}
f(\boldsymbol {s_{B_i}}|  s_{a_i-1},s_{b_i+1}, \mu, \phi, \sigma) &= 
\prod_{t=0}^{c_i-1}
f(s_{a_i+t}|  \boldsymbol {s_{a_i-1:a_i+t-1  }},s_{b_i+1}, \mu, \phi, \sigma)\\
&= 
\prod_{t=0}^{c_i-1}
f(s_{a_i+t}|  s_{a_i+t-1},s_{b_i+1}, \mu, \phi, \sigma),
\end{split}
\end{equation*}
where  $f(s_{a_i+t}|  s_{a_i+t-1},s_{b_i+1}, \mu, \phi, \sigma)$ is the univariate normal density with mean
\begin{equation*}
\mu_{a_i+t|a_i+t-1,b_i+1} = \frac{1}{1-\phi^{2(c_i+1-t))}} \left( (\phi-\phi^{2c_i+1-2t}) (s_{a_i+t-1}-\mu) + (\phi^{c_i-t} - \phi^{c_i+2-t} (s_{b_i+1}-\mu)) \right),
\end{equation*}
and variance
\begin{equation*}
\sigma^2_{a_i+t|a_i+t-1,b_i+1} = \frac{\sigma^2}{1-\phi^2}\left(1 - \frac{1}{1-\phi^{2(c_i+1-t)}} \left(\phi^2 -2\phi^{2(c_i-t+1)}+\phi^{2(c_i-t)} \right)\right).
\end{equation*}

So we can sample $\boldsymbol {s_{B_i}} = (s_{a_i+t})_{t=0, \ldots, s_i-1}$ conditioned on $s_{a_i-1},s_{b_i+1}, \mu, \phi, \sigma$ recursively by
\begin{equation*}
\begin{split}
&s_{a_i+t} \sim N(\mu_{a_i+t|a_i+t-1,b_i+1}, \sigma^2_{a_i+t|a_i+t-1,b_i+1}),
\end{split}
\end{equation*}
for $t = 0, \ldots, c_i-1$ and then $\boldsymbol {s_{B_i}} - \boldsymbol{\mu_{B_i|}}$ is a sample from $N(\boldsymbol 0,\Sigma_{B_i|})$.

\subsection*{Sampling of the constant parameters in the ancillary augmentation}
\label{sec:AA}
To sample $\mu, \phi$ and $\sigma$ in (AA) we deploy an adaptive random walk Metropolis-Hastings scheme as suggested by \cite{garthwaite2016adaptive}, where tuning parameters are selected automatically using the Robbins Monro process (\cite{robbins1985stochastic}). For sampling, it is convenient to move to unconstrained parameter spaces which is achieved by the following transformations
\begin{equation*}
\psi = \ln(\sigma), ~~
\xi = F_Z(\phi).
\end{equation*}
Here $F_Z(x) = \frac{1}{2} \log(\frac{1+x}{1-x})$ is Fisher's Z transformation. The from \eqref{eq:priors} implied log prior densities for $\xi$ and $\psi$ are given by
\begin{equation*}
\begin{split}
\ln(\pi(\xi)) &= (a_{\phi}-1)\ln(F_Z^{-1}(\xi)+1)+(b_{\phi}-1) \ln(1-F_Z^{-1}(\xi))+\ln(1-(F_Z^{-1}(\xi))^2) + c_1\\
\ln(\pi(\psi)) &= -\psi - \frac{1}{2B_{\sigma}}\exp(2\psi) + 2\psi + c_2.
\end{split},
\end{equation*}

where $c_1 \in \mathbb{R}$ and $c_2 \in \mathbb{R}$ are constants. The log posterior density in (AA) is obtained as
\begin{equation*}
\begin{split}
lp_{(AA)}(\mu, & \xi, \psi, s_0, \boldsymbol{\tilde s_{1:T}}|Y) = \sum_{t=1}^T \ln(f(\boldsymbol {y_t}|s_t(\boldsymbol{ \tilde s_{1:T}}, \mu, F_Z^{-1}(\xi), \exp(\psi))))- \frac{1}{2} \sum_{t=1}^T \tilde s_t^2 \\
&+ \ln(\varphi\left(s_0|\mu, \frac{\exp(\psi)^2}{1-F_Z^{-1}(\xi)^2}\right))+\ln(\pi(\mu)) + \ln(\pi(\xi)) + \ln(\pi(\psi)) + c_3,
\end{split}
\end{equation*}
where $c_3 \in \mathbb{R}$ is a constant. We sample $(\mu, \phi, \sigma)$ in two blocks, one block for $\mu$ and one block for $(\phi, \sigma)$.

\subsubsection*{Update for $\mu$}
To sample the mean parameter $\mu$ from its full conditional we propose a new state $\mu_{prop}$ in the $r$-th iteration of the MCMC procedure by
\begin{equation*}
\mu_{prop} \sim N(\mu_{cur}, \sigma_{MH,\mu}^{r-1}), 
\end{equation*}
where $\mu_{cur}$ is the current value for $\mu$. The proposal $\mu_{prop}$ is accepted with probability
\begin{equation*}
R = \exp(lp_{(AA)}(\mu_{prop}, \xi, \psi, s_0, \boldsymbol{\tilde s_{1:T}}|Y)-lp_{(AA)}(\mu_{cur}, \xi, \psi, s_0, \boldsymbol{\tilde s_{1:T}}|Y))
\end{equation*}
and the scaling parameter $\sigma_{MH,\mu}^r$ is updated according to \cite{garthwaite2016adaptive} by
\begin{equation*}
\ln(\sigma_{MH,\mu}^{r}) = \ln(\sigma_{MH,\mu}^{r-1}) + 4.058\frac{(R - 0.44)}{r-1}.
\end{equation*}
The scaling parameter is reduced, if the acceptance probability is larger than 0.44 and increased if the acceptance probability is smaller than 0.44. We target an average acceptance probability of 0.44, as recommended by \cite{roberts2001optimal} for univariate random walk Metropolis-Hastings. The constant $4.058$ controls the step size and is chosen as suggested by \cite{garthwaite2016adaptive}. 

\subsubsection*{Joint update for $\phi$ and $\sigma$}

In the $r$-th iteration, a two dimensional proposal $(\xi_{prop}, \psi_{prop})$ for $(\xi_{}, \psi_{})$   is obtained by
\begin{equation*}
(\xi_{prop}, \psi_{prop})^\top \sim N_2( (\xi_{cur}, \psi_{cur})^\top, \Sigma_{MH,\xi,\psi}^{r-1} ),
\end{equation*}
where $(\xi_{cur}, \psi_{cur})$ are the current values. The proposal is accepted with probability
\begin{equation*}
R = \exp(lp_{(AA)}(\mu, \xi_{prop}, \psi_{prop}, s_0, \boldsymbol{\tilde s_{1:T}}|Y)-lp_{(AA)}(\mu, \xi_{cur}, \psi_{cur}, s_0, \boldsymbol{\tilde s_{1:T}}|Y)).
\end{equation*}

For adapting the covariance matrix we follow a suggestion of \cite{garthwaite2016adaptive}. Let $I_n$ denote the $n$-dimensional identity matrix. We set $\Sigma_{MH,\xi,s}^{r} = I_2$ if $r<100$ and 
\begin{equation*}
\Sigma_{MH,\xi,s}^{r} = (\sigma_{MH,\xi,s}^{r})^2\left(\hat \Sigma^{r} + \frac{(\sigma_{MH,\xi,s}^{r})^2}{r} I_2\right)    ~~~~~~ \text{if } r \geq 100.
\end{equation*}
Here $\hat \Sigma^{r}$ is the empirical covariance matrix of $(\xi^i,\psi^i)_{i=1, \ldots r}$, the first $r$ samples for $(\xi,\psi)$, and

\begin{equation*}
\ln(\sigma_{MH,\xi,s}^{r}) = \ln(\sigma_{MH,\xi,s}^{r-1}) + 6.534\frac{(R - 0.234)}{r-1}.
\end{equation*} 

The matrix $\hat \Sigma^{r} + \frac{(\sigma_{MH,\xi,\psi}^{r})^2}{r} I_2$ is a positive definite estimate of the covariance matrix. This covariance estimate is scaled by $(\sigma_{MH,\xi,\psi}^{r})^2$ to obtain the covariance matrix for the proposal in the next iteration. The scaling $(\sigma_{MH,\xi,\psi}^{r})^2$ is tuned to achieve an average acceptance probability of 0.234 as suggested by \cite{roberts1997weak} for multivariate random walk Metropolis-Hastings.
To reduce computational cost the empirical covariance matrix $\hat\Sigma^{r}$ can be updated in every step by the following recursion (see e.g. \cite{bennett2009numerically})
\begin{equation*}
\hat\Sigma^{r} = \frac{r-2}{r-1} \hat\Sigma^{r-1} + \frac{1}{r} ((\xi^{r},\psi^{r})^\top - \hat{\boldsymbol\mu}^{r-1})((\xi^{r},\psi^{r})^\top - \hat{\boldsymbol\mu}^{r-1})^\top,
\end{equation*}
where $\hat{\boldsymbol\mu}^{r-1}$ is the sample mean of $(\xi^i, \psi^i)_{i=1, \ldots, r-1}$. We also update the sample mean recursively by
\begin{equation*}
\hat{\boldsymbol\mu}^{r}=\frac{1}{r}((r-1)\hat{\boldsymbol\mu}^{r-1} + (\xi^{r},\psi^{r})^\top).
\end{equation*}

We have seen that the adaptions for the $\mu$ and the $(\phi,\sigma$) updates tend to be very small after burn-in and therefore we only adapt during the burn-in period. This also ensures a correct sampling procedure without the need to verify the validity of an adaptive MCMC scheme.

\section*{Appendix B. The standardized skew Student t distribution}

According to \cite{azzalini2003distributions}, the density of the univariate skew Student t distribution with parameters $\xi \in \mathbb{R}, \omega \in (0,\infty), \alpha \in\mathbb{R}$ and $df \in (0,\infty)$   is given by
\begin{equation*}
st(x|\xi, \omega, \alpha, df )= \frac{2}{\omega}t(x |df)T\left(\alpha \frac{x-\xi}{\omega} \sqrt{\frac{df +1}{\left(\frac{x-\xi}{\omega}\right)^2+df}}\vast|df+1\right), 
\end{equation*}
where $t(\cdot|df)$ is the density function of the univariate Student t distribution with $df$ degrees of freedom and $T(\cdot|df)$ the corresponding distribution function.
The expectation and variance of a random variable $X$ following a skew Student t distribution with parameters $\xi, \omega, \alpha$ as above and $df > 2$ are given by 
\begin{equation*}
E(X) = \xi +  \omega b_{df} \delta, \text{ and } Var(X) = \omega^2\left(\frac{df}{df -2 } - b_{df}^2 \delta^2  \right),
\end{equation*}
where $\delta = \frac{\alpha^2}{\sqrt{1+ \alpha^2}}$ and $b_{df}  = \sqrt{\frac{df}{\pi}} \frac{\Gamma(\frac{df-1}{2})}{\Gamma(\frac{df}{2}) }$.
If we set
\begin{equation*}
\begin{split}
\omega &= \sqrt{\frac{1}{\left(\frac{df}{df -2 } - b_{df}^2 \delta^2  \right)}} ~\text{ and }~
\xi = - \omega b_{df} \delta = - \sqrt{\frac{1}{\left(\frac{df}{df -2 } - b_{df}^2 \delta^2  \right)}} b_{df} \delta,
\end{split}
\end{equation*}
only the parameters $\alpha$ and $df$ remain unknown and the random variable has zero mean and a variance of one. We refer to the corresponding distribution as the standardized skew Student t distribution. Its density is denoted by $sst$ and is obtained as
\begin{equation}
sst(x|\alpha, df) = st\left(x \svast|  - \sqrt{\frac{1}{\left(\frac{df}{df -2 } - b_{df}^2 \delta^2  \right)}} b_{df} \delta, \sqrt{\frac{1}{\left(\frac{df}{df -2 } - b_{df}^2 \delta^2  \right)}}, \alpha, df\right).
\label{eq:sst}
\end{equation}

\section*{Appendix C. Additional material for the bivariate dynamic mixture copula (Section \ref{sec:application})}

\begin{figure}[H]
\centerline{%
\includegraphics[trim={0 7cm 0 0},width=1.0\textwidth]{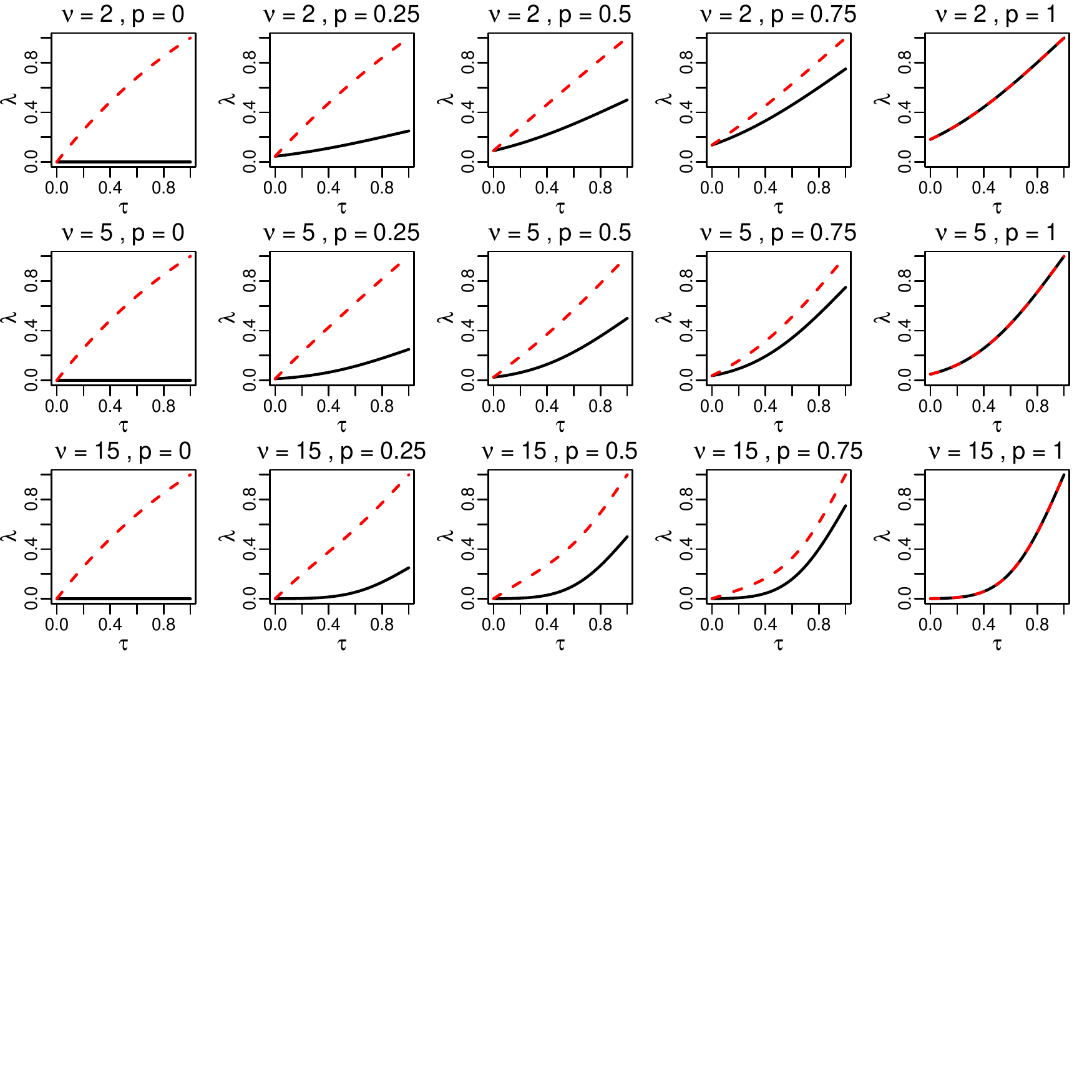}%
}%
\caption{Upper (red, dashed) and lower (black) tail dependence coefficient of the mixture copula defined in \eqref{eq:mixcop} plotted against Kendall's $\tau$ for different values of $\nu$ and $p$.}
\label{fig:taildep}
\end{figure}

\begin{figure}[H]
\centerline{%
\includegraphics[trim={0 10.5cm 0 0},width=1.0\textwidth]{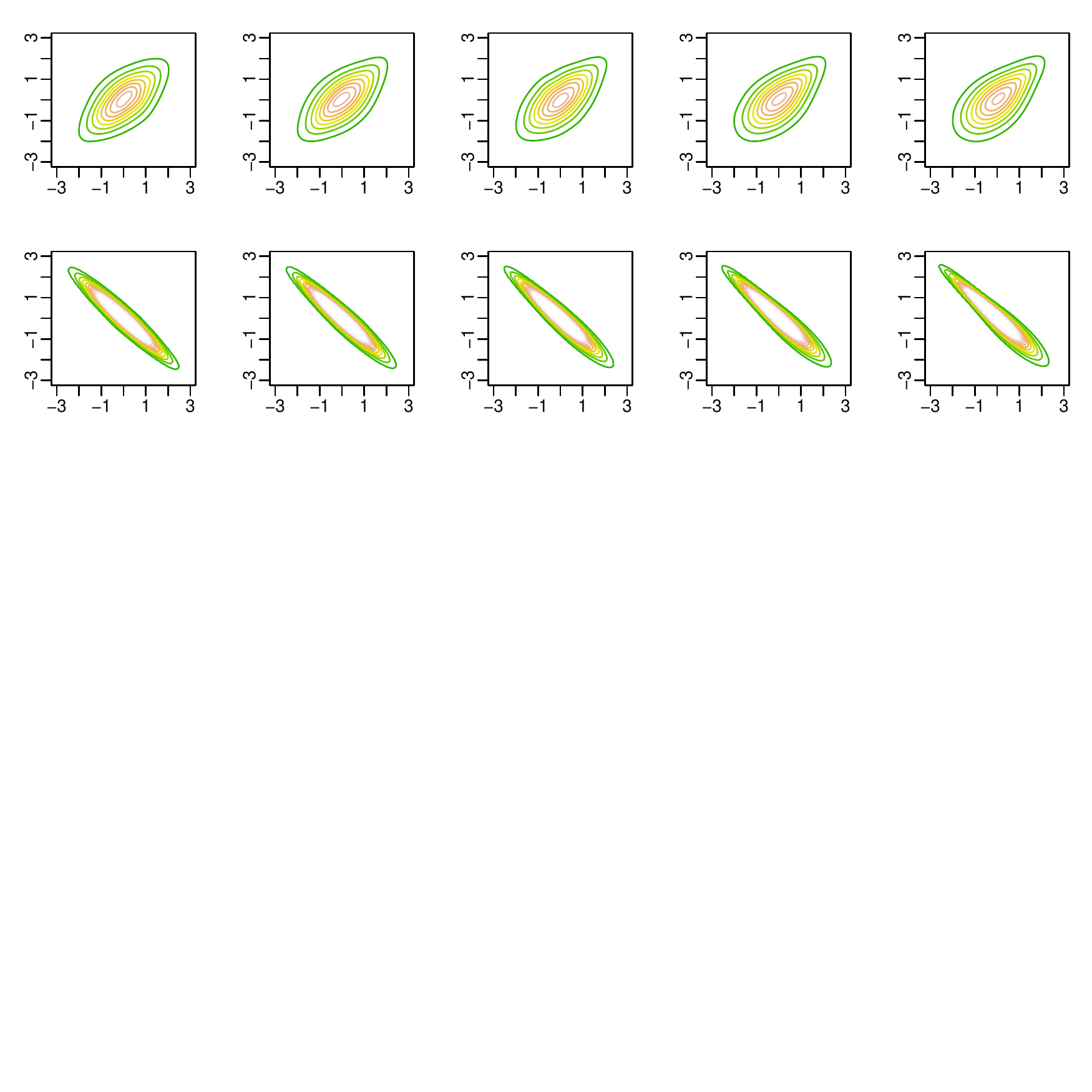}%
}%
\caption{Normalized contour plots for the mixture copula model in \eqref{eq:mixcop} with $\tau=0.4$ (top row), $\tau=-0.8$ (bottom row), $\nu=5$ and $p=1,0.75,0.5,0.25,0$ (from left to right).}
\label{fig:mix_cont}
\end{figure}

\section*{Appendix D. Additional material for the application (Section \ref{sec:application})}

\label{append:app}

\subsection*{Daily log returns}
\begin{figure}[H]
\centerline{%
\includegraphics[trim={0 11cm 0 0},width=1.0\textwidth]{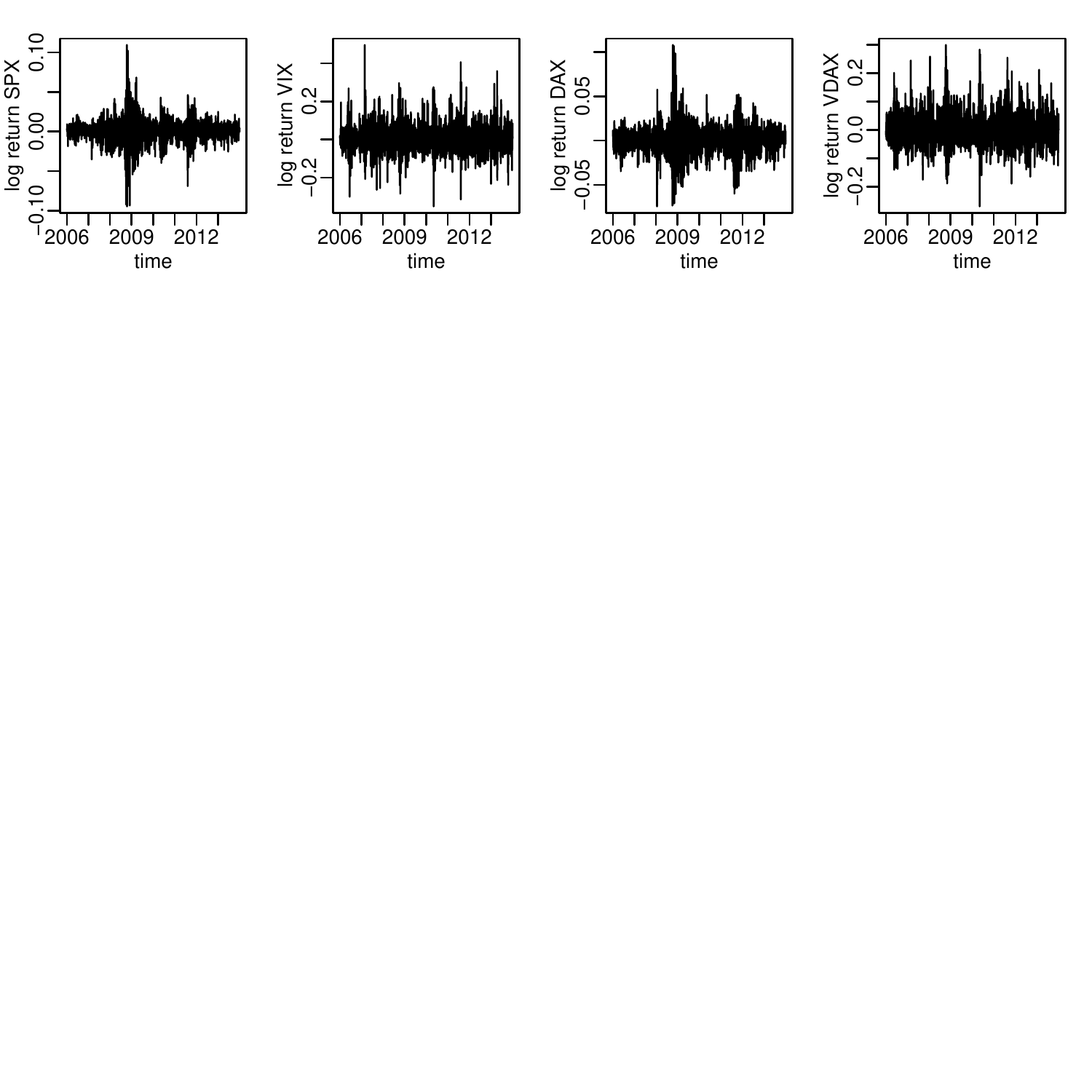}%
}%
\caption{Daily log returns of the four indices SPX, VIX, DAX, VDAX from 2006 to 2013 plotted against time.}
\label{fig:dlogret}
\end{figure}

\subsection*{Posterior statistics}

\begin{table}[H]
\centering
\begin{tabular}{crrrr}
  \hline
& mode & $5\%$ quantile & $95\%$ quantile & effective sample size \\ 
   \hline  
   \multicolumn{5}{c}{\textbf{SPX}}  \\
$\mu$ & -9.32 & -9.94 & -8.70 & 13896.29 \\ 
  $\phi$ & 0.99 & 0.98 & 1.00 & 565.10 \\ 
  $\sigma$ & 0.15 & 0.13 & 0.19 & 208.31 \\ 
  $\alpha$ & -0.51 & -0.80 & -0.22 & 4293.41 \\ 
  $df$ & 6.84 & 5.46 & 10.40 & 1821.93 \\ 
      \hline
   \multicolumn{5}{c}{\textbf{VIX}}  \\
 $\mu$ & -5.65 & -5.80 & -5.50 & 3586.38 \\ 
 $\phi$ & 0.90 & 0.84 & 0.93 & 362.48 \\ 
  $\sigma$ & 0.36 & 0.28 & 0.48 & 311.76 \\ 
  $\alpha$ & 1.33 & 0.97 & 1.73 & 1376.24 \\ 
   $df$& 9.30 & 6.50 & 15.14 & 1251.32 \\ 
      \hline
   \multicolumn{5}{c}{\textbf{DAX}}  \\ 
  $\mu$ & -8.89 & -9.30 & -8.50 & 19573.46 \\ 
  $\phi$ & 0.99 & 0.97 & 0.99 & 598.27 \\ 
  $\sigma$ & 0.15 & 0.12 & 0.19 & 249.07 \\ 
  $\alpha$ & -0.48 & -0.80 & -0.06 & 5662.26 \\ 
  $df$& 9.74 & 7.31 & 15.11 & 2483.44 \\ 
      \hline
  \multicolumn{5}{c}{\textbf{VDAX}}  \\
  $\mu$ & -6.06 & -6.21 & -5.89 & 8086.52 \\ 
  $\phi$ & 0.96 & 0.92 & 0.97 & 416.02 \\ 
  $\sigma$ & 0.18 & 0.13 & 0.24 & 293.01 \\ 
  $\alpha$ & 0.96 & 0.66 & 1.27 & 3305.73 \\ 
  $df$& 8.35 & 6.44 & 12.88 & 1386.65 \\ 
   \hline
\end{tabular}
\caption{Posterior mode estimates, posterior quantiles and effective sample sizes for the univariate skew Student t stochastic volatility models for the four indices SPX, VIX, DAX, VDAX.}
\label{tab:marginal}
\end{table}

\begin{table}[H]
\centering
\begin{tabular}{crrrr}
  \hline
& mode & $5\%$ quantile & $95\%$ quantile & effective sample size \\ 
   \hline
   \multicolumn{5}{c}{\textbf{(SPX,VIX)}}  \\
$\mu$ & -0.74 & -0.77 & -0.71 & 1862.77 \\ 
  $\phi$ & 0.94 & 0.85 & 0.97 & 306.33 \\ 
  $\sigma$ & 0.05 & 0.03 & 0.08 & 215.92 \\ 
  $p$ & 0.29 & 0.13 & 0.44 & 1436.78 \\ 
  $\nu$ & 9.03 & 5.29 & 41.14 & 1039.38 \\ 
  \hline   
   \multicolumn{5}{c}{\textbf{(DAX,VDAX)}}  \\
$\mu$ & -0.81 & -0.84 & -0.78 & 1785.30 \\ 
  $\phi$ & 0.86 & 0.73 & 0.92 & 285.29 \\ 
  $\sigma$ & 0.10 & 0.06 & 0.13 & 207.29 \\ 
  $p$ & 0.66 & 0.50 & 0.81 & 1406.11 \\ 
  $\nu$ & 8.30 & 5.91 & 34.32 & 756.50 \\  
 \hline
\end{tabular}
\caption{Posterior mode estimates, posterior quantiles and effective sample sizes for the dynamic mixture copula models for the pairs (SPX,VIX) and (DAX,VDAX).}
\label{tab:cop}
\end{table}

\subsection*{Calculating the log predictive score}
We describe in detail how we proceed for model $\mathcal{M}_{dyn}^{mix}$. We consider $T+K$ observations of dimension two, stored in the data matrix $Y_{1:(T+K);1:2}$, where the first $T$ observations are used to train the model and the last $K$ are used for evaluation. 
\vspace*{0.5cm}

\noindent
\textbf{Step 1: (Model fitting based on the training period)}
\begin{itemize}
\item We fit two marginal skew Student t stochastic volatility models to $\boldsymbol {y_{1:T;1}}$ and $\boldsymbol {y_{1:T;2}}$. This yields $R_{train}$ draws of the parameters denoted by $\boldsymbol {s_{1:T;j}^{st,r}}$, $\mu_{j}^{st,r}$, $\phi_{j}^{st,r}$, $\sigma_{j}^{st,r}$, $\alpha_{j}^{st,r}$ and $df_{j}^{st,r}$, $r=1, \ldots, R_{train}$ and corresponding posterior mode estimates  $\boldsymbol{\hat s_{1:T;j}^{st}}$, $\hat\mu_{j}^{st}$, $\hat\phi_{j}^{st}$, $\hat\sigma_{j}^{st}$, $\hat\alpha_{j}^{st}$ and $\hat{df}_{j}^{st}$ for $j=1,2$.
 
\item We estimate the copula data 
\begin{equation*}
\hat u_{tj} = ssT\left(y_{tj}\exp(-\frac{\hat s_{tj}^{st}}{2})\Big | \hat \alpha_j^{st}, \hat df_j^{st}\right) 
\end{equation*}
 for $t=1, \ldots, T, j=1,2$.

\item We fit the dynamic bivariate mixture copula model introduced in \eqref{eq:dynmix} based on the pseudo copula data $\hat U_{1:T;1:2}$ and obtain posterior draws $\boldsymbol {s_{1:T}^{cop,r}}$, $\mu^{cop,r}$, $\phi^{cop,r}$, $\sigma^{cop,r}$, $\nu^{cop,r}$, $p^{cop,r}$ for $r=1, \ldots, R_{train}$ and corresponding posterior mode estimates $\boldsymbol{ \hat s_{1:T}^{cop}}$, $\hat\mu^{cop}$, $\hat\phi^{cop}$, $\hat\sigma^{cop}$, $\hat\nu^{cop}$, $\hat p^{cop}$.
\end{itemize}

\noindent
\textbf{Step 2: (The one-day ahead predictive density)}
\vspace*{0.2cm}

Estimating the one-day ahead predictive density at time $T+k, 1\leq k\leq K$ would usually require to fit daily models with observations up to time $T+k-1$ for $k=1, \ldots, K$. In order to save computational resources we use another approach where we only update the dynamic parameters, i.e. the log variances and Kendall's $\tau$. For the constant parameters we use the estimates from the training period $1, \ldots, T$. In this case we found that it is enough to only consider a time horizon of 100 time points, i.e. to estimate a dynamic parameter at time $T+k$ we consider data in the period $T+k-100, \ldots, T+k-1$.  We proceed as follows  to obtain the one-day ahead predictive density at time point $T+k$ with $1\leq k\leq K$.
\begin{itemize}
\item We consider a skew Student t stochastic volatility model as in \eqref{eq:stsv}, where we keep the parameters $\mu$, $\phi$, $\sigma$, $\alpha$ and $df$ fixed and only update the latent log variances. Therefore we draw the latent log variances  $\boldsymbol {s_{(T+k-100):(T+k-1);j}^{st}}$ conditional on $\boldsymbol {y_{(T+k-100):(T+k-1);j}}$, $ \hat\mu_{j}^{st}$, $ \hat\phi_{j}^{st}$, $ \hat\sigma_{j}^{st}$, $\hat{ df}_{j}^{st}$ and $ \hat\alpha_{j}^{st}$ for $j=1,2$. We denote the draws by $\boldsymbol {s^{st,r}_{(T+k-100):(T+k-1);j}}$, $r=1, \ldots, R_{test}$ for $j=1,2$. Corresponding posterior mode estimates are denoted by $\boldsymbol{ \hat s_{(T+k-100):(T+k-1);j}^{st}}, j=1,2$.
\item We estimate the copula data via the probability integral transform, i.e.
for $j=1,2$ and $t = T+k-100, \ldots T+k-1$ we calculate
\begin{equation*}
\hat u_{tj}=ssT\left(y_{tj}\exp(-\frac{\hat s_{tj}^{st}}{2}) \Bigg | \hat \alpha_{j}^{st}, \hat {df}_{j}^{st}\right).
\end{equation*}
\item We fit the dynamic mixture copula model to the pseudo copula data $\hat U_{(T+k-100):(T+k-1);(1:2)}$ where we keep the constant parameters fixed. We only update $\boldsymbol {s_{(T+k-100):(T+k-1)}^{cop}}$ conditional on 
$\hat U_{(T+k-100):(T+k-1);(1:2)}$, $\hat\mu^{cop}$, $\hat\phi^{cop}$, $\hat\sigma^{cop}$, $\hat \nu^{cop}$, $\hat p^{cop}$. The corresponding draws are denoted by $\boldsymbol {s^{cop,r}_{(T+k-100):(T+k-1)}}, r=1, \ldots, R_{test}$ and the posterior mode estimates by $\boldsymbol{ \hat s_{(T+k-100):(T+k-1)}^{cop}}$.

\item For $j=1,2$, we obtain an estimate for the log variance at time point $T+k$ as $\hat s_{T+k;j}^{st} = \hat\mu_{j}^{st} + \hat\phi_{j}^{st} (\hat s^{st}_{T+k-1;j} - \hat\mu_{j}^{st})$.  

\item We obtain an estimate for Fisher's Z transform of Kendall's $\tau$ at time point $T+k$, $ $ as  $\hat s_{T+k}^{cop} = \hat\mu^{cop} + \hat\phi^{cop} (\hat s_{T+k-1}^{cop} - \hat \mu^{cop})$.  

\item The predictive density evaluated at $(y_1,y_2)$ is given by 
\begin{equation*}
f_{T+k}^{p}(y_1,y_2)=c_{T+k}^{p}(y_1,y_2) g_{T+k}^{p}(y_1,y_2),
\end{equation*}
with
\begin{equation*}
c_{T+k}^{p}(y_1,y_2)=c^M\left(ssT\left(x_1 \Big|\hat \alpha_{1}^{st}, \hat{df}_{1}^{st}\right), ssT\left(x_2\Big|\hat \alpha_{2}^{st}, \hat{df}_{2}^{st}\right);F_Z^{-1}(\hat s_{T+k}^{cop}), \hat \nu^{cop}, \hat p^{cop}     \right),
\end{equation*}
where $c^M$ is the density of the mixture copula defined in \eqref{eq:mixcop} and
\begin{equation*}
g_{T+k}^{p}(y_1,y_2)=sst\left(x_1 \Big |\hat \alpha_{1}^{st}, \hat{df}_{1}^{st}\right) sst\left(x_2\Big |\hat \alpha_{2}^{st}, \hat{df}_{2}^{st}\right) \exp(-\frac{\hat s_{T+k;1}^{st}}{2}) \exp(-\frac{\hat s_{T+k;2}^{st}}{2}),
\end{equation*}
with $x_j = {y_j}{\exp(-\hat s_{T+kj}^{st}/2)}$ for $j=1,2$. 
\end{itemize} 

\noindent
\textbf{Step 3: (The cumulative pseudo log predictive score)}
\vspace*{0.2cm}

The cumulative pseudo log predictive score is obtained as
\begin{equation*}
\begin{split}
LP &= \sum_{k=1}^K\log(f^p_{T+k}(y_{T+k;1},y_{T+k;2})).\\
\end{split}
\end{equation*}

During the training period we run $R_{train} = 31000$ iterations with a burn-in of 1000, while for updating only the dynamic parameters 11000 iterations with a burn-in of 1000 is enough, i.e. we use $R_{test}=11000$.

\newpage
\noindent
{\Huge\bf Supplementary material}

\setcounter{section}{0}

\section{Elliptical slice sampling}
We assume that the posterior density for a parameter vector $\boldsymbol \theta$ given data $D$ is proportional to
\begin{equation}
f(\boldsymbol \theta|D) \propto \ell(\boldsymbol \theta|D) \varphi(\boldsymbol \theta | \boldsymbol 0 , \Sigma),
\label{eq:postdens}
\end{equation} 
where $\ell(\boldsymbol \theta|D)$ is the likelihood function and $\varphi(\boldsymbol \theta | \boldsymbol 0 , \Sigma)$ is the multivariate normal density with zero mean and covariance matrix $\Sigma$. 
\cite{murray2010elliptical} consider the Metropolis-Hastings sampler of \cite{bernardo1998regression} where a proposal $\boldsymbol \theta'$ is obtained from the following stochastic representation
\begin{equation}
\boldsymbol\theta' = \sqrt{1 - \alpha^2} \boldsymbol\theta + \alpha \boldsymbol v, \boldsymbol v \sim N(\boldsymbol 0,\Sigma).
\label{eq:hellipse}
\end{equation}
Here $\alpha \in [-1,1]$ is a fixed step size parameter. The proposal is accepted with probability 
\begin{equation}
\min\left(1,\frac{\ell(\boldsymbol\theta')}{\ell(\boldsymbol\theta)}\right).
\label{eq:acc}
\end{equation}
Elliptical slice sampling adapts the step size parameter $\alpha$ during sampling. This eliminates the need to select the parameter before sampling and it may be a better approach for situations where good choices of the step size parameter depend on the region of the state space. 
\cite{murray2010elliptical} first suggest alternatively to propose a new state by
\begin{equation}
\boldsymbol \theta' = \cos(\omega)\boldsymbol\theta  + \sin(\omega) \boldsymbol v, \boldsymbol v \sim N(\boldsymbol 0,\Sigma).
\label{eq:fellipse}
\end{equation}
Here the angle $\omega$ corresponds to the step size. As we move $\omega$ towards zero the proposal gets closer to the initial value $\boldsymbol \theta$.  
\cite{murray2010elliptical} argue that \eqref{eq:fellipse} provides a more flexible choice for the proposals compared to $\eqref{eq:hellipse}$, if the parameter $\omega$ is also updated, which is here the case. In elliptical slice sampling we first draw an angle $\omega$ from the uniform distribution on $[0,2\pi]$ and obtain a proposal as outlined in \eqref{eq:fellipse}. This proposal is accepted according to \eqref{eq:acc}. If the proposal is not accepted a new angle is selected with a slice sampling approach (\cite{neal2003slice}) such that the angle approaches zero as more samples are rejected. This ensures that at some point the proposal will be accepted. The approach is outlined in  Algorithm \ref{algo}. 
\cite{murray2010elliptical} show that this samples from a Markov chain, where \eqref{eq:postdens} is the corresponding stationary distribution.

\begin{algorithm}[H]
\caption{Elliptical slice sampling}
\label{algo}
\begin{algorithmic}[1]
\STATE $\boldsymbol v \sim N(\boldsymbol 0,\Sigma)$
\STATE $ u\sim uniform(0,1) $ 
\STATE $\omega \sim uniform(0,2\pi)$
\STATE $\omega_{min}= \omega - 2\pi,~~ \omega_{max} = \omega$
\STATE $\boldsymbol\theta' =  \cos(\omega) \boldsymbol\theta +  \sin(\omega) \boldsymbol v$

\WHILE{$\frac{l(\boldsymbol\theta')}{l(\boldsymbol\theta)} \leq u$}
\IF{$\omega < 0 $} \STATE $\omega_{min} = \omega $
\ELSE \STATE $\omega_{max} = \omega $
  \ENDIF
  \STATE $\omega \sim uniform(\omega_{min},\omega_{max})$
\STATE $\boldsymbol\theta' = \cos(\omega) \boldsymbol\theta  + \sin(w)\boldsymbol v $
\ENDWHILE
\end{algorithmic}
\end{algorithm}

\section{The DCC-GARCH model}
The DCC-GARCH model was introduced by \cite{engle2002dynamic}. Mathematical propoerties were developed by \cite{engle2001theoretical}. In the DCC-GARCH model  it is assumed that a $d$ dimensional random vector $\boldsymbol {\epsilon_t} \in \mathbb{R}^d$ at time $t$ is multivariate normal distributed with zero mean and dynamic covariance matrix $H_t \in \mathbb{R}^{d \times d}$, i.e. 
\begin{equation*}
\boldsymbol {\epsilon_t}|H_t \sim N(\boldsymbol 0,H_t),
\end{equation*}
for $t=1, \ldots, T$. The covariance matrix can be written as 
\begin{equation*}
H_t = D_tC_tD_t, 
\end{equation*}
where  $C_t$ is a $d \times d$ correlation matrix and $D_t$ is a $d \times d$ diagonal matrix. The diagonal matrix $D_t$ contains the marginal standard deviations. We assume GARCH($P_j$,$Q_j$) innovations for the $j$-th diagonal entry of $D_t$, i.e.
\begin{equation*}
d_{t,j}^2 = \omega_j  + \sum_{p=1}^{P_j}\alpha_{pj} \epsilon_{t-p,j}^2 +  \sum_{q=1}^{Q_j}\beta_{qj} d_{t-q,j}^2, 
\end{equation*}
where
\begin{itemize}
\item $\omega_j > 0$,
\item $d_{t,0}>0$,
\item  $\sum_{p=1}^{P_i} \alpha_{jp} + \sum_{q=1}^{Q_i} \beta_{qj}<1$ and the roots of $1 - \sum_{p=1}^{P_i} \alpha_{jp}Z^p + \sum_{q=1}^{Q_i} \beta_{qj}Z^q$ lie outside the unit circle,
\item $\alpha_{jp}$ for all $p \in \{1, \ldots P_j\}$ and $\beta_{jp}$ for all $q \in \{1, \ldots Q_j\}$  are such that $d_{tj}^2$ is positive. 
\end{itemize} 
  The correlation matrix is decomposed as follows
\begin{equation*}
C_t = diag(L_t)^{-1/2}L_tdiag(L_t)^{-1/2}, 
\end{equation*}
where $L_t$ is a positive definite matrix. Further we denote by
 \begin{equation*}
\boldsymbol r_t = D_t^{-1} \boldsymbol {\epsilon_t}
\end{equation*}
 the standardized return and obtain $\bar L$ as
\begin{equation*}
\bar L = \frac{1}{T}\sum_{t=1}^T \boldsymbol{r_t} \boldsymbol{r_t^{\top}}.
\end{equation*} 
 For $L_t$ we assume the following dynamic structure
\begin{equation*}
L_t = \left(1 - \sum_{m=1}^M a_m - \sum_{n=1}^{N}b_n\right) \bar L + \sum_{m=1}^M a_m \boldsymbol {r_{t-m}} \boldsymbol {r_{t-m}^{\top}} + \sum_{n=1}^N b_n L_{t-n},
\end{equation*}
where 
\begin{itemize}
\item $\alpha_m \geq 0$ for all $m \in \{1, \ldots M\}$ and $b_n \geq 0 $ for all $n \in \{1, \ldots, N\}$,
\item $\sum_{m=1}^M a_m + \sum_{n=1}^N b_n < 1$,
\item $L_0$ is positive definite.
\end{itemize}
\cite{engle2001theoretical} show that under the above conditions $H_t$ is a proper covariance matrix. The DCC(M,N)-GARCH(P,Q) model is obtained by setting $P_j = P$ and $Q_j=Q$ for $j = 1, \ldots, d$.

\section{Further results for the application}
\subsection*{Results for the marginal models}
\begin{figure}[H]
\centerline{%
\includegraphics[trim={0 3.5cm 0 0},width=1.0\textwidth]{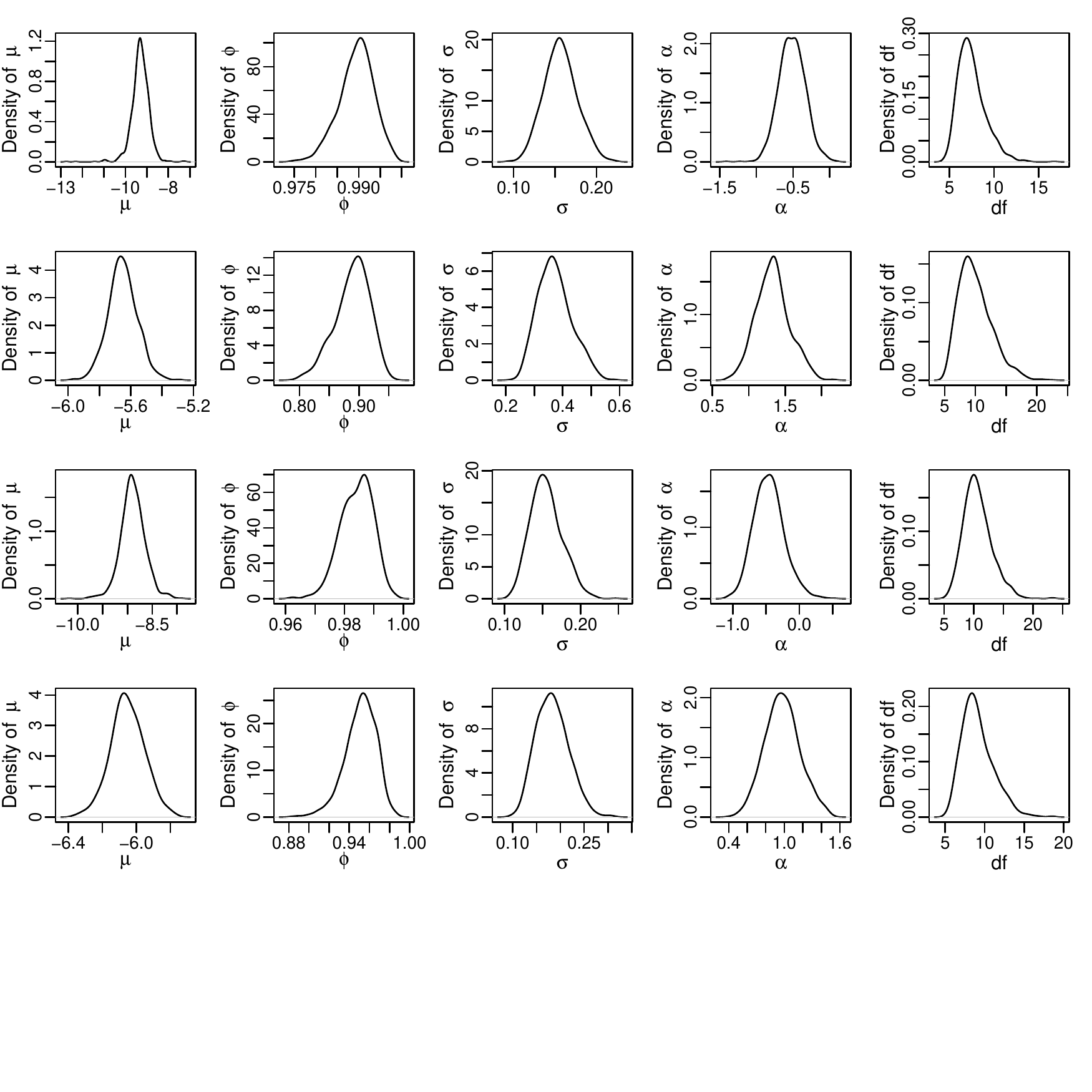}%
}%
\caption{Estimated posterior densities based on 30000 MCMC iterations after a burn-in of 1000 for the parameters of the univariate skew Student t stochastic volatility models for SPX, VIX, DAX and VDAX (from top to bottom row).}
\label{fig:marg_dens}
\end{figure}

\begin{figure}[H]
\centerline{%
\includegraphics[trim={0  3.25cm 0 0},width=1.0\textwidth]{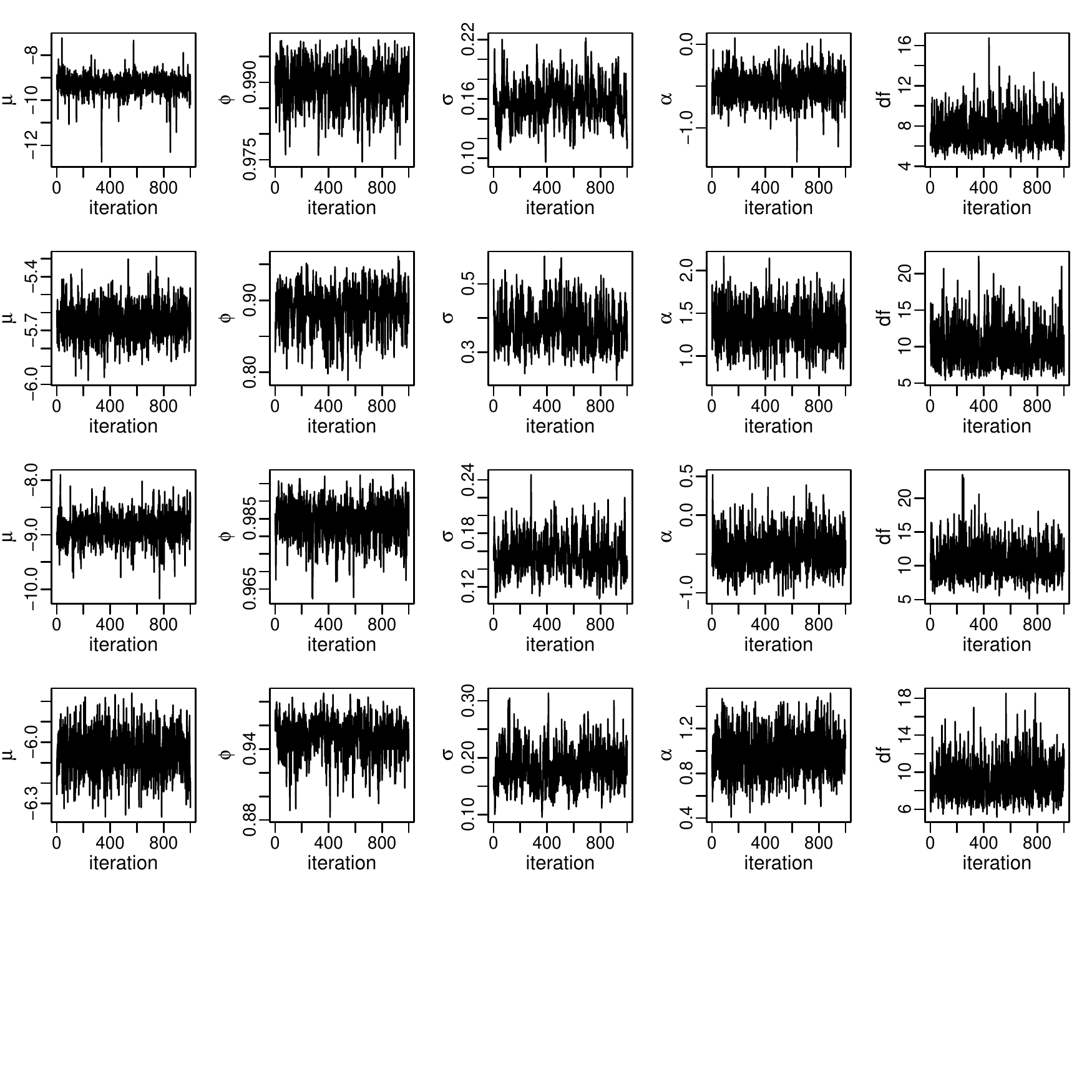}%
}%
\caption{Trace plots of 1000 MCMC draws based on a total of 31000 iterations, where the first 1000 draws are discarded for burn-in and the remaining 30000 draws are thinned with factor 30. The trace plots are shown for parameters of the univariate skew Student t stochastic volatility models for SPX, VIX, DAX and VDAX (from top to bottom row).}
\label{fig:svs_trace_application}
\end{figure}

\subsection*{Results for the dependence models}

\begin{figure}[H]
\centerline{%
\includegraphics[trim={0  10.75cm 0 0},width=1.0\textwidth]{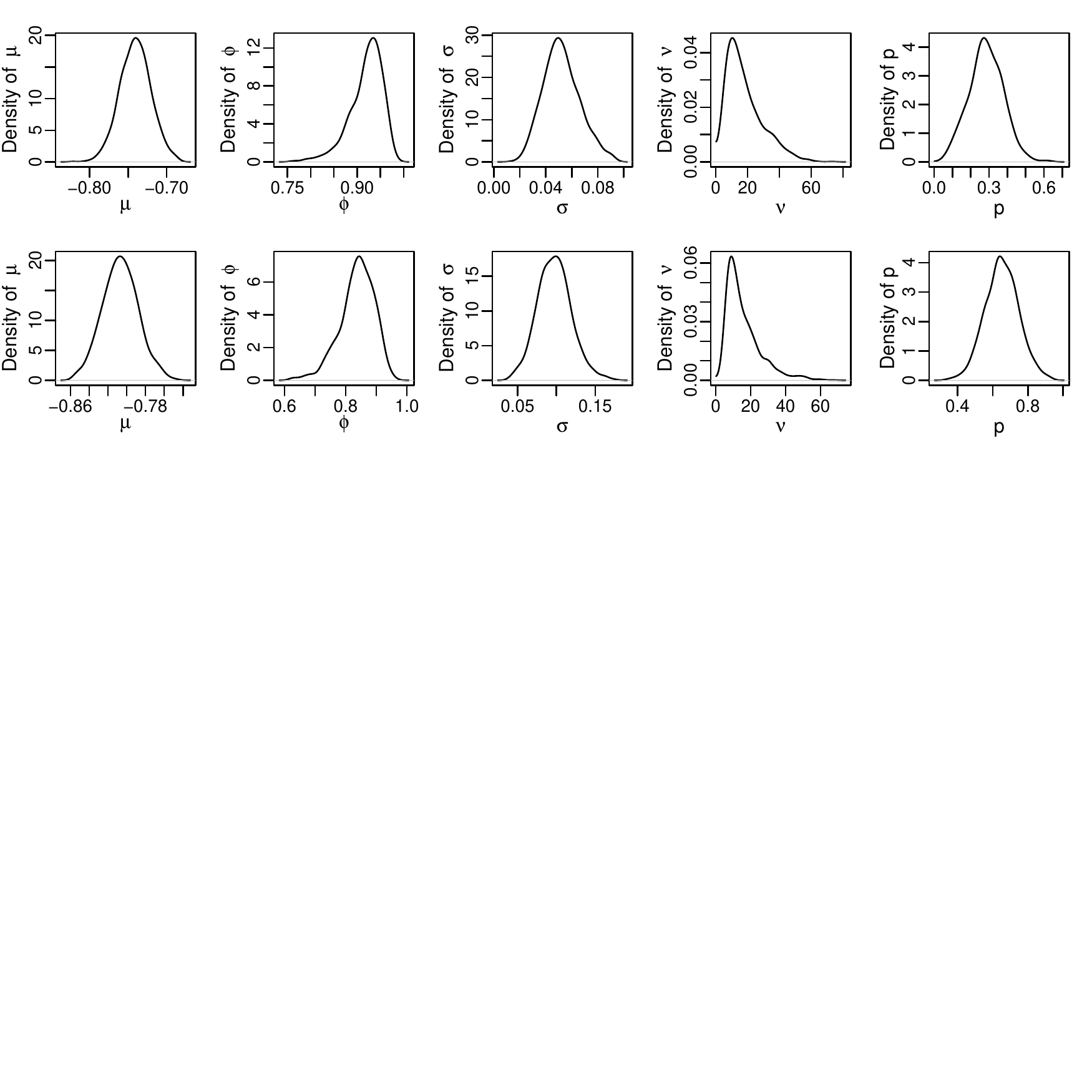}%
}%
\caption{Estimated posterior densities based on 30000 MCMC iterations after a burn-in of 1000 for parameters of the dynamic mixture copula model for (SPX,VIX) in the top row and for (DAX,VDAX) in the bottom row.}
\label{fig:mix_density_application}
\end{figure}

\begin{figure}[H]
\centerline{%
\includegraphics[trim={0  10.5cm 0 0},width=1.0\textwidth]{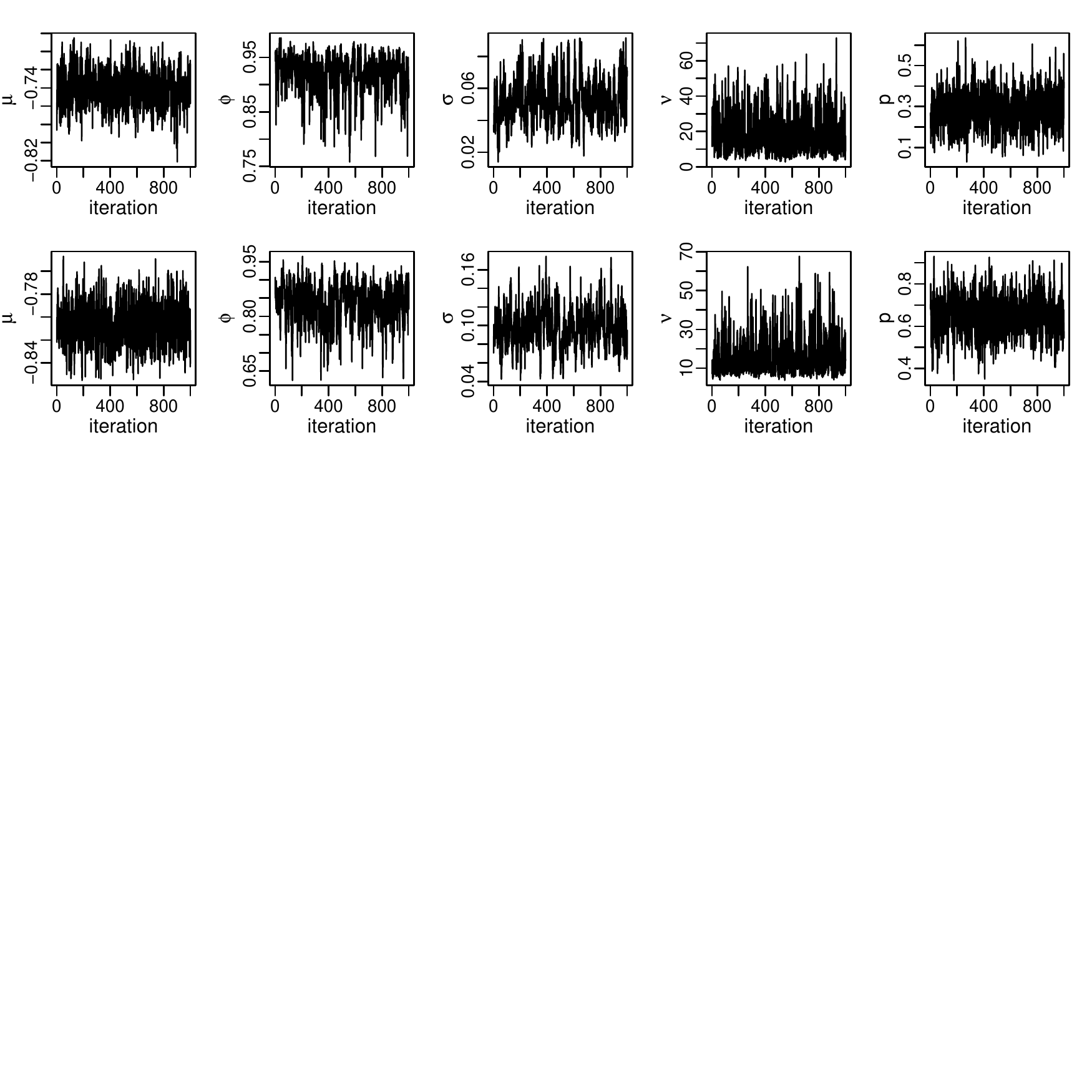}%
}%
\caption{Trace plots of 1000 MCMC draws based on a total of 31000 iterations, where the first 1000 draws are discarded for burn-in and the remaining 30000 draws are thinned with factor 30. The trace plots are shown for parameters of the dynamic mixture copula model for (SPX,VIX) in the top row and for (DAX,VDAX) in the bottom row.}
\label{fig:mix_trace_application}
\end{figure}

\end{document}